\pdfoutput=1

\documentclass[11pt,twoside,a4paper,cmspaper,final,collab]{cms-tdr}
\def\svnVersion{439062P}\def\cmsCernNoTag{CERN-EP-2017-074}\def\cmsCernDate{\today}\def\cmsMessage{Published in Physics Letters B as \href{http://dx.doi.org/10.1016/j.physletb.2017.09.053}{\doi{10.1016/j.physletb.2017.09.053.}}}

\begin{document}\cmsNoteHeader{EXO-15-007}

\hyphenation{had-ron-i-za-tion}
\hyphenation{cal-or-i-me-ter}
\hyphenation{de-vices}
\RCS$Revision: 439035 $
\RCS$HeadURL: svn+ssh://svn.cern.ch/reps/tdr2/papers/EXO-15-007/trunk/EXO-15-007.tex $
\RCS$Id: EXO-15-007.tex 439035 2017-12-15 18:25:30Z glandsbe $
\newlength\cmsFigWidth
\ifthenelse{\boolean{cms@external}}{\setlength\cmsFigWidth{0.49\textwidth}}{\setlength\cmsFigWidth{0.75\textwidth}}
\ifthenelse{\boolean{cms@external}}{\providecommand{\cmsLeft}{top\xspace}}{\providecommand{\cmsLeft}{left\xspace}}
\ifthenelse{\boolean{cms@external}}{\providecommand{\cmsRight}{bottom\xspace}}{\providecommand{\cmsRight}{right\xspace}}
\cmsNoteHeader{EXO-15-007}
\newcommand{\st}{\ensuremath{S_{\mathrm{T}}}\xspace}
\newcommand{\mbh}{\ensuremath{M_{\mathrm{BH}}}\xspace}
\newcommand{\mbhmin}{\ensuremath{M_{\mathrm{BH}}^\text{min}}\xspace}
\newcommand{\MPl}{\ensuremath{M_\mathrm{Pl}}\xspace}
\renewcommand{\MD}{\ensuremath{M_\mathrm{D}}\xspace}
\newcommand{\MS}{\ensuremath{M_\mathrm{S}}\xspace}
\newcommand{\gs}{\ensuremath{g_\mathrm{S}}\xspace}
\newcommand{\photonjet}{\ensuremath{\gamma\text{+jets}}\xspace}
\newcommand{\W}{\PW\xspace}
\providecommand{\NA}{\ensuremath{\text{---}}\xspace}
\providecommand{\BLACKMAX} {{\textsc{BlackMax}}\xspace}
\ifthenelse{\boolean{cms@external}}{\newcommand{\cmsTable}[1]{#1}}{\newcommand{\cmsTable}[1]{\resizebox{\textwidth}{!}{#1}}}

\title{Search for black holes and other new phenomena in high-multiplicity final states in proton-proton collisions at $\sqrt{s}= 13\TeV$}

\date{\today}

\abstract{A search for new physics in energetic, high-multiplicity final states has been performed using proton-proton collision data collected with the CMS detector at a center-of-mass energy of 13\TeV and corresponding to an integrated luminosity of 2.3\fbinv. The standard model background, dominated by multijet production, is determined exclusively from control regions in data. No statistically significant excess of events is observed. Model-independent limits on the product of the cross section and the acceptance of a new physics signal in these final states are set and further interpreted in terms of limits on the production of black holes. Semiclassical black holes and string balls with masses as high as 9.5\TeV, and quantum black holes with masses as high as 9.0\TeV are excluded by this search in the context of models with extra dimensions, thus significantly extending limits set at a center-of-mass energy of 8\TeV with the LHC Run 1 data.}

\hypersetup{%
pdfauthor={CMS Collaboration},%
pdftitle={Search for black holes and other new phenomena in high-multiplicity final states in proton-proton collisions at sqrt(s) = 13 TeV},%
pdfsubject={CMS},%
pdfkeywords={CMS, physics, black holes}}

\maketitle

\section{Introduction\label{sec:introduction}}

The standard model (SM)~\cite{SM1,SM2,SM3} of particle physics is a remarkably successful theory. However, several outstanding problems remain. One of these is the ``hierarchy problem"~\cite{hierarchy}, i.e., the vast separation between the electroweak energy scale and the scale at which gravity becomes strong. The latter, referred to as the ``Planck scale" (\MPl), is some 17 orders of magnitude greater than the former. There are various theoretical extensions of the SM that address the hierarchy problem, such as supersymmetry (SUSY) and models with extra dimensions.

In many of these models, high-multiplicity, energetic final states naturally occur. Strong single or pair production of various new physics signals result in multijet final states, often accompanied by energetic leptons and/or invisible particles resulting in transverse momentum (\pt) imbalance in the event. Examples include a large variety of SUSY signals, both with $R$-parity~\cite{R-parity} conservation~\cite{SUSYprimer} and violation~\cite{RPVModel}, and signals associated with technicolor models~\cite{TC}, axigluons~\cite{Axigluons}, colorons~\cite{Colorons0,Colorons1,Colorons2,Colorons3}, and various models with low-scale gravity.

In this Letter, we describe a model-independent search for new physics in high-multiplicity final states, and explicitly test the predictions of two possible solutions to the hierarchy problem. One of these solutions invokes a model with $n$ large extra dimensions, colloquially known as the ``ADD model", named after its proponents, Arkani-Hamed, Dimopoulos, and Dvali~\cite{add1,add2,add3}. The other solution is based on the Randall--Sundrum model~\cite{RS1, RS2}, called the ``RS1 model". In this model a single, compact extra spatial dimension is warped, and the SM particles are localized on a TeV-scale brane, while gravity originates on the second, Planck brane, separated from the TeV brane in the extra dimension.

In the ADD model, the fundamental multidimensional Planck scale (\MD) is related to the ``apparent" 3-dimensional Planck scale \MPl as:
\begin{equation}
\MD = \frac{1}{r}\left(\frac{r\MPl}{\sqrt{8\pi}}\right)^\frac{2}{n+2},
\end{equation}
where $r$ is the compactification radius or the characteristic size of extra dimensions.

In the RS1 model, the analog of the ADD scale \MD is defined as a function of the exponential warp factor $k$ and the compactification radius $r$:
\begin{equation}
\MD = \frac{\MPl}{\sqrt{8\pi}}\re^{-\pi k r}.
\end{equation}
In both models, $\MD$ can be of order 1\TeV, thus eliminating the hierarchy of scales and alleviating the hierarchy problem.

At high-energy hadron colliders, such as the CERN LHC, if the collision energy exceeds \MD, both the ADD and RS1 models allow for the formation of microscopic black holes (BHs)~\cite{dl,gt,RSBH1,RSBH2,RSBH3}. In the simplest scenario, microscopic BHs are produced when the distance of closest approach between two colliding particles is less than the Schwarzschild radius $R_\mathrm{S}$, which for a BH in a ($3+n$)-dimensional space is given by~\cite{Myers}:
\begin{equation}
    R_\mathrm{S}= \frac{1}{\sqrt{\pi}\MD} \left[ \frac{\mbh}{\MD} \left(\frac{8\Gamma(\frac{n+3}{2})}{n+2}\right)\right]^{\frac{1}{n+1}},
    \label{eq:schw}
\end{equation}
where \mbh is the mass of the BH. The parton-level production cross section of such processes is expected to be simply $\pi R_\mathrm{S}^2$~\cite{dl,gt}. In more complicated scenarios with energy loss during the formation of the BH horizon, the production cross section could significantly depart from this simple geometrical formula. Since the production of BHs is a threshold phenomenon, we assume a minimum mass threshold $\mbhmin \geq \MD$.

In the semiclassical case, corresponding to $\mbh \gg \MD$, BHs evaporate rapidly via Hawking radiation~\cite{Hawking}, with a lifetime of order $10^{-27}$ seconds. As gravity couples universally to the energy-momentum tensor and does not distinguish between various particle species, microscopic BHs decay democratically into all SM degrees of freedom, i.e., all SM particle species with all possible values of quantum numbers, such as spin, color, and charge. The final state is therefore populated by a variety of energetic particles, such as hadrons (jets), leptons, photons, and neutrinos. Due to the large number of color degrees of freedom, about 75\% of the particles produced are expected to be quarks and gluons. The final state may contain significant transverse momentum imbalance from the presence of neutrinos, which constitute about 5\% of the decay products. Other processes, such as the decay of W and Z bosons, or of heavy-flavor quarks, and the possible emission of gravitons or the formation of a noninteracting stable BH remnant, contribute to the transverse momentum imbalance as well.

The semiclassical approximation breaks down when the mass of the BH approaches \MD and the BH becomes a quantum object or a quantum black hole (QBH). These objects do not obey the usual BH thermodynamics and hence decay much more rapidly than their semiclassical counterparts. Their decays are characterized by the presence of only a few particles, e.g., a pair of jets~\cite{QBH1,QBH2,QBH3}. These QBHs could also decay into lepton flavor violating final states, as preserving baryon number or lepton numbers separately is typically not a requirement of the decay process~\cite{QBH1,QBH2}.

In addition to semiclassical BHs and QBHs, one could also explore stringy precursors of BHs, called ``string balls" (SBs)~\cite{sb}. Such objects, which arise in string theory, are highly excited, long, folded strings that form below the BH production threshold. Like semiclassical BHs, SBs evaporate thermally, but at a constant Hagedorn temperature~\cite{Hagedorn} independent of the SB mass, and also produce a large number of energetic particles in the final state, with the composition similar to that for a semiclassical BH.  String balls undergo a phase transition into ordinary semiclassical BHs when their mass reaches $\MS/\gs^2$~\cite{sb}, where $\MS$ and $\gs$ are the string scale and the string coupling constant, respectively. For an SB mass between $\MS/\gs$ and the BH transition, $\MS/\gs^2$, the parton-level cross section saturates at $\sigma \sim 1/\MS^2$, while for lighter SBs, it grows as $\sigma \sim \gs^2 M_\mathrm{SB}^2/\MS^4$~\cite{sb}.

While our choice of final states is inspired by the production of microscopic BHs, in this Letter we focus on a generic search (Section~\ref{sec:limits}) that can be used to probe a large class of new-physics models. Consequently, our emphasis is on the exploration of a multiparticle final state with a model-independent search.

During Run 1 of the LHC, a number of searches for semiclassical and quantum BHs were performed at a center-of-mass energy of 8\TeV. A review of these results can be found in Ref.~\cite{GL-Springer}. The limits on the minimum BH mass set by these searches lie in the 6\TeV range. With the increased LHC center-of-mass energy of 13\TeV, the BH phase space can be probed much more extensively, as was demonstrated in the recent ATLAS publications~\cite{ATLASjj,ATLAS-inclusive13,ATLAS-QBH13}, which set BH mass limits reaching  9\TeV.

\section{Analysis strategy}

The most challenging aspect of the analysis presented in this Letter is accurately describing the QCD multijet background, since the BH signal leads to a broad excess in the \st spectrum, rather than a narrow peak. Here, \st is defined as the scalar sum of the transverse energies of jets, leptons, photons, as well as the missing transverse energy (\MET, defined as the magnitude of the transverse momentum imbalance in an event, as detailed in Section~\ref{sec:event_reconstruction}):
\begin{equation}
    \st = \left(\sum_{i=1}^N E_{{\rm T},i}\right) + E_{\rm T}^{\rm miss},
                    \label{eq:ST}
\end{equation}
where $N$ is the total number of final-state objects (excluding the \MET), or the object multiplicity. For the QCD background, the final-state objects are almost exclusively jets and the \MET is expected to be small, so the \st variable is reduced to a scalar sum of the transverse momenta of the jets. The signal region for this search typically lies in the high-multiplicity regime where the QCD multijet background is dominated by higher-order effects. These effects have not been calculated for high-multiplicity final states, and therefore an accurate simulation of the QCD multijet background, pertinent to our signal region, does not yet exist.

This significant hurdle is mitigated by predicting the QCD multijet background directly from data using a new technique developed in Run 1 of the LHC~\cite{CMSBH1,CMSBH2,CMSBH3}. Studies performed with simulated QCD multijet events and with data at low object multiplicities show that the shape of the \st distribution above its turn-on threshold is approximately independent of the multiplicity of the final state. This observation is consistent with the development of the parton shower via nearly collinear emission, which approximately conserves the \st value, up to the effects of additional jets falling below the kinematic threshold. For this reason one can predict the \st spectrum of a multijet final state using samples of dijet or trijet events. This feature provides a powerful tool to predict the shape of the \st spectrum at higher multiplicity using a low-multiplicity control region. The method has found wide applicability in various CMS searches, such as a search for stealth SUSY~\cite{stealth} and a search for multijet resonances~\cite{EXO-13-001}. An earlier CMS analysis~\cite{CMSBH1} also considered other kinematic variables, such as the invariant mass or transverse invariant mass of the event. However, the multiplicity invariance is not exhibited by these variable to the degree shown by the \st variable.

In this Letter we follow closely the methodology of Refs.~\cite{CMSBH1,CMSBH2,CMSBH3} geared toward a multiparticle final state, dominated by QCD multijets in the case of semiclassical BHs, and toward a dijet final state for the QBHs. The variable \st is the single discriminating variable used in the analysis, chosen for its robustness against variations in the BH evaporation model and its lack of sensitivity to the relative abundance of various particles produced. This variable encompasses the total transverse energy in an event and is therefore useful in discriminating between the signal and the background. There is a minimum transverse energy ($\ET$) threshold of 50\GeV~\cite{CMSBH1} that each of the objects (including the \MET) has to satisfy to be counted toward the definition of \st.  The exact choice of the \et threshold is not particularly important; the 50\GeV threshold is chosen as it makes the analysis insensitive to additional interactions in the same or adjacent bunch crossings (pileup) and moderates the effect of initial-state radiation, which generally spoils the \st invariance.

The \st distributions are produced in a number of inclusive object multiplicity bins ($N \ge N^\text{min}$). The background is estimated exclusively from collision data via the appropriately chosen control regions. This approach does not rely on the Monte Carlo (MC) description of the backgrounds. In addition, this method has the advantage of being more sensitive and less model dependent than exclusive searches in specific final states, e.g., the lepton+jets final state~\cite{ATLAS-ljets1,ATLAS-ljets2}. The ATLAS Collaboration has recently adopted a similar inclusive search strategy based on the 2012 $\sqrt{s} = 8$\TeV~\cite{ATLAS-inclusive} and the 2015 $\sqrt{s} = 13$\TeV data~\cite{ATLAS-inclusive13}.

\section{The CMS detector}
\label{cms_detector}

The central feature of the CMS apparatus is a superconducting solenoid of 6\unit{m} internal diameter, providing a magnetic field of 3.8\unit{T}. Within the solenoid volume are a silicon pixel and strip tracker, a lead tungstate crystal electromagnetic calorimeter (ECAL), and a brass and scintillator hadron calorimeter (HCAL), each composed of a barrel and two endcap sections. Forward calorimeters extend the pseudorapidity coverage provided by the barrel and endcap detectors up to $|\eta| < 5$. Muons are measured in gas-ionization detectors embedded in the steel flux-return yoke outside the solenoid.

The silicon tracker measures charged particles within the pseudorapidity range $\abs{\eta} < 2.5$. It consists of 1440 silicon pixel and 15\,148 silicon strip detector modules. For nonisolated particles of $1 < \pt < 10\GeV$ and $\abs{\eta} < 1.4$, the track resolutions are typically 1.5\% in \pt and 25--90 (45--150)\mum in the transverse (longitudinal) impact parameter \cite{TRK-11-001}

In the region $\abs{\eta} < 1.74$, the HCAL cells have widths of 0.087 in pseudorapidity and 0.087 radians in azimuth ($\phi$). In the $\eta$-$\phi$ plane, and for $\abs{\eta} < 1.48$, the HCAL cells map on to $5{\times}5$ arrays of ECAL crystals to form calorimeter towers projecting radially outwards from close to the nominal interaction point. For $\abs{\eta } > 1.74$, the coverage of the towers increases progressively to a maximum of 0.174 in $\Delta \eta$ and $\Delta \phi$. Within each tower, the energy deposits in ECAL and HCAL cells are summed to define the calorimeter tower energies, subsequently used to provide the energies and directions of hadronic jets.

The first level of the CMS trigger system~\cite{TRG-12-001}, composed of custom hardware processors, combines information from the calorimeters and muon detectors to select the most interesting events in a fixed time interval of 3.2\mus. The high-level trigger (HLT) processor farm further decreases the event rate from around 100\unit{kHz} to less than 1\unit{kHz}, before data storage.

A detailed description of the CMS detector, together with a definition of the coordinate system used and the relevant kinematic variables, can be found in Ref.~\cite{Chatrchyan:2008aa}.

\section{Event reconstruction}
\label{sec:event_reconstruction}

The analysis is based on proton-proton collision data corresponding to an integrated luminosity of 2.3\fbinv, collected with the CMS detector in 2015. The trigger chosen for the analysis is based on the total transverse energy of jets in an event. At the HLT, events are selected if they passed an 800\GeV threshold on the scalar sum of the transverse momenta of all hadronic jets, which are reconstructed with the particle-flow (PF) algorithm~\cite{PFT-paper}, as described below. The trigger is nearly 100\% efficient for \st above 1\TeV.

In the subsequent analysis, we select events with at least one reconstructed vertex~\cite{TRK-11-001} within 24 (2) cm of the nominal interaction point measured parallel (perpendicular) to the LHC beamline. The vertex with the highest sum of the transverse momenta squared of the associated tracks is chosen as the hard-scattering vertex.

The reconstruction of physics objects in the event is based on the PF algorithm that identifies each single particle in an event (photon, electron, muon, charged hadron, neutral hadron) using an optimized combination of all subdetector information. The energy of photons is directly obtained from the ECAL measurement, corrected for zero-suppression effects~\cite{gammaPerformance}. The energy of electrons is determined from a combination of the track momentum at the main interaction vertex, the corresponding ECAL cluster energy, and the energy sum of all bremsstrahlung photons attached to the track. The energy of muons is obtained from the corresponding track momentum. The energy of charged hadrons is determined from a combination of the track momentum and the corresponding ECAL and HCAL energy, corrected for zero-suppression effects and for the response function of the calorimeters to hadronic showers. The energy of neutral hadrons is obtained from the corresponding corrected ECAL and HCAL energy. Finally, the \MET is defined as the absolute value of the vectorial \pt sum of all the PF candidates reconstructed in an event~\cite{Khachatryan:2014gga}.

For each event, hadronic jets are clustered from the PF candidates using the anti-\kt algorithm~\cite{Cacciari:2008gp} with a distance parameter of 0.4, as implemented in the {\FASTJET} package~\cite{FastJet}. The jet momentum is determined as the vectorial sum of the momenta of all its constituents, and is found to be within 5 to 10\% of the true (particle-level) momentum over the whole \pt spectrum and the detector acceptance. Jet energy corrections are derived from simulation and in situ measurements of the energy balance in dijet, multijet, $\gamma$+jet, and leptonic Z+jet events~\cite{Chatrchyan:2011ds,Khachatryan:2016kdb}. The contribution of charged hadrons that do not originate from the hard-scattering vertex (``pileup"), but are clustered during the reconstruction of a jet, is subtracted. Corrections based on the jet area~\cite{rho-method} are applied to remove the energy contribution of neutral hadrons from pileup interactions. The jet energy resolution (JER) is approximately 8\% at 100\GeV and 4\% at 1\TeV~\cite{Chatrchyan:2011ds,Khachatryan:2016kdb}. The minimum threshold on the corrected $\ET$ of the jets used in the analysis is 50\GeV, and jets are accepted in the full pseudorapidity range of the CMS calorimeters ($\abs{\eta} < 5$). To reduce contamination from poorly reconstructed muons, the fraction of the jet momentum carried by a muon is required to be less than 80\%. Also, to suppress jets due to rare, spurious anomalous calorimeter signals, jets must contain at least two particles, one of which is a charged hadron, and the jet energy fraction carried by neutral PF candidates (neutral hadrons and photons) should be less than 90\%. These criteria have an efficiency greater than 99\% per jet.

Details of muon reconstruction can be found in Ref.~\cite{MUO-10-004}. Muon candidates are required to satisfy a minimum \pt threshold of 50\GeV and to be within $\abs{\eta} < 2.4$. The transverse impact parameter and the longitudinal distance of the track associated with the muon with respect to the primary vertex is required to be less than 2 and 5\unit{mm}, respectively, to reduce contamination from cosmic muons. The muon candidate is required to have at least one energy deposit in the pixel tracker and at least six deposits in the silicon strip tracker. The global track fit to the tracker trajectory and to at least two muon detector segments must have a $\chi^2$ per degree of freedom of less than 10.

Details of electron reconstruction can be found in Ref.~\cite{EGM-13-001}. Electron candidates are required to have $\ET> 50\GeV$, $\abs{\eta} < 2.5$, excluding the $1.44 < \abs{\eta} < 1.57$ transition region between the ECAL barrel and endcap detectors where the reconstruction is suboptimal, and to pass standard identification criteria, corresponding to a working point with an average efficiency of 80\% per electron.

Both muons and electrons are required to be isolated from other energy deposits in the tracker and the calorimeters. The relative isolation $\mathcal{I}$  is defined as the ratio of the sum of transverse energies of photons, and charged and neutral hadrons in a cone of radius of 0.4\,(0.3) in the $\eta$-$\phi$ space centered on the muon (electron) candidate to the \pt of the lepton:
\begin{equation}
{\cal I} = \frac{(\sum_i  E_{\rm T}^i) - E_{\rm T}^{\rm PU}}{\pt^\ell},
\end{equation}
where the sum runs over charged hadrons originating from the hard-scattering vertex, neutral hadrons, and photons. The numerator of the ratio is corrected for contributions due to pileup ($\ET^\mathrm{PU}$), using the fraction of energy carried by the charged hadrons originating from other vertices to estimate the contribution of neutral particles from pileup for muons, and an average area method~\cite{rho-method}, as estimated with {\FASTJET} for electrons. Muons (electrons) are required to have relative isolation values less than 0.15 (0.10).

Photons are required to have $\pt > 50$\GeV and to be in the same fiducial region as the electrons, and are reconstructed with a standard algorithm~\cite{gammaPerformance} using a medium working point. They are also required to be isolated, with the definition of isolation tuned to yield constant efficiency as a function of photon $\pt$. The pileup-corrected charged-hadron transverse energy sum within the isolation cone of radius of 0.3 is required to be less than 1.37 (1.10)\GeV in the barrel (endcaps). A similarly defined neutral-hadron isolation sum is required to be less than $1.06\GeV + 0.014\pt + 0.000019\GeV^{-1} \pt^2$ in the barrel and $2.69\GeV + 0.0139\pt + 0.000025\GeV^{-1}\pt^2$ in the endcaps. Finally, the pileup-corrected isolation sum of any additional photon candidates in the isolation cone must be less than $0.28\GeV + 0.0053\pt$ ($0.39\GeV + 0.0034\pt$) in the barrel (endcaps).

\section{Signal simulation}
\label{sec:signal}

Signal events are simulated using dedicated MC event generators. For semiclassical BHs, the {\BLACKMAX} v2.02.0~\cite{BlackMax} and {\CHARYBDIS2} v1.003~\cite{Charybdis2} generators are used, as summarized in Table~\ref{table:signal} and detailed below. Quantum BHs are generated using the \textsc{qbh} v3.00 generator~\cite{QBH3}. The fragmentation and hadronization of parton-level signal samples is done with {\PYTHIA 8.205}~\cite{pythia8}, with the underlying event tune CUETP8M1~\cite{CUET}.

\begin{table*}[htb]
\topcaption{Generator settings for various semiclassical BH model points probed in this analysis. These parameters are defined in Refs.~\protect\cite{BlackMax} and \protect\cite{Charybdis2} for the  {\BLACKMAX}  and {\CHARYBDIS2} generators, respectively. The generator settings not specified are kept at their default values.}
\label{table:signal}
\centering
\cmsTable{\begin{tabular}{cccccccc}
\multicolumn{8}{c}{\BLACKMAX}\\
\hline
Model &  \multicolumn{2}{c}{Choose\_a\_case}	& \multicolumn{2}{c}{Mass\_loss\_factor}	& \multicolumn{2}{c}{Momentum\_loss\_factor}	& \multicolumn{1}{c}{turn\_on\_graviton}	\\\hline
B1	&     \multicolumn{2}{c}{tensionless\_nonrotating}	&      \multicolumn{2}{c}{0}		&	\multicolumn{2}{c}{0}		&  \multicolumn{1}{c}{FALSE} \\
B2 	&     \multicolumn{2}{c}{rotating\_nonsplit}	& 	\multicolumn{2}{c}{0}		& 	\multicolumn{2}{c}{0}		&  \multicolumn{1}{c}{FALSE} \\
B3 	&     \multicolumn{2}{c}{rotating\_nonsplit}	&     \multicolumn{2}{c}{0.1}		&  \multicolumn{2}{c}{0.1}		&  \multicolumn{1}{c}{TRUE} \\
\hline\\[1ex]
\multicolumn{8}{c}{\CHARYBDIS2}\\
\hline
 Model 	& 	\textsc{bhspin} 	&\textsc{mjlost} 	& \textsc{yrcsc} 		& \textsc{nbodyaverage}	& \textsc{nbodyphase} 	& \textsc{nbodyvar}	& \textsc{rmstab} \\
 \hline
C1    	&      TRUE     	& FALSE		& FALSE			& FALSE			& TRUE			&	TRUE	& FALSE\\
C2		&	FALSE	& FALSE		& FALSE			& FALSE			& TRUE			&	TRUE 	& FALSE\\	
C3		&	TRUE	& FALSE		& FALSE			& TRUE			& FALSE			& 	FALSE 	& FALSE\\
C4		&	TRUE	& TRUE		& TRUE			& FALSE			& TRUE			& 	TRUE 	& FALSE\\
C5		& 	TRUE	& TRUE		& TRUE			& FALSE			& FALSE			& 	FALSE	& TRUE\\
\end{tabular}}
\end{table*}

In the semiclassical case, several models are explored. We simulate the following scenarios: nonrotating BHs (models B1, C2), rotating BHs without energy loss (models B2, C1) and with an alternative evaporation model (C3), rotating BHs with 10\% loss of mass and angular momentum  (B3), rotating BHs with Yoshino--Rychkov bounds~\cite{YR1} (model C4), and rotating BHs with a stable remnant with the mass equal to the multidimensional Planck scale \MD (model C5, for which additionally the {\sc Charybdis2} parameter NBODY was changed from its default value of 2 to 0). The generator parameters used for semiclassical BH signal models are given in Table~\ref{table:signal}.

The parameters associated with the BH signal generation, as can be inferred from Eq.~(\ref{eq:schw}), are the number of extra dimensions $n$, the multidimensional Planck scale \MD, and the mass of the BH \mbh. For the semiclassical case, $n= 2$, 4, and 6, while \MD is varied between 2 and 9\TeV, and \mbh is varied in the range between \MD and 11\TeV. In the case of the QBHs, $n= 1$--6, while $\MD =2$--9\TeV and $\mbh \geq \MD$ with the upper bound kept at 11\TeV. For QBHs in the ADD model we use $n \ge 2$, while in the RS1 model $n$ is restricted to 1.

String balls are generated using the {\CHARYBDIS2} event generator for the nonrotating scenario. The number of extra dimensions is fixed at $n=6$, as it was in earlier CMS publications. Note also that as the dependence of the SB dynamics on $n$ is only implicit, via the relationship between the Planck and string scales. The mass of the string ball ($M_\mathrm{SB}$) is varied between 5 and 10\TeV. Four different benchmark scenarios are considered:
\begin{itemize}
	\item $\MD = 5.93\TeV$, $\MS$ = 1.1\TeV, $\gs= 0.2$;
	\item $\MD = 5.36\TeV$, $\MS$ = 1.1\TeV, $\gs= 0.3$;
	\item $\MD = 6.80\TeV$, $\MS$ = 1.5\TeV, $\gs= 0.4$;
	\item $\MD = 8.57\TeV$, $\MS$ = 2.0\TeV, $\gs= 0.5$.
\end{itemize}
These benchmarks are chosen such that the various regimes of the SB dynamics are probed, mainly between the first transition at $M_\mathrm{SB} = \MS/\gs$ and the BH transition at $M_\mathrm{SB} = \MS/\gs^2$. Above the second transition, SBs turn into semiclassical BHs, so the standard BH analysis fully applies. (We note that a significant fraction of the SB parameter space probed by the ATLAS Collaboration~\cite{ATLAS-inclusive13} in fact falls into the BH, and not the SB regime.) In all cases the fundamental Planck scale \MD is chosen to satisfy the matching condition at the SB/BH transition point: $\sigma_\mathrm{BH}(\MS/\gs^2) = \sigma_\mathrm{SB}(\MS/\gs^2)  = 1/\MS^2$, where $\sigma_\mathrm{BH}(\mbh)$ and $\sigma_\mathrm{SB}(M_\mathrm{SB})$ are the BH and SB parton-level production cross sections, as functions of their masses, respectively.

The choice of parton distribution functions (PDFs) used in this analysis was made as a result of detailed studies of the PDF dependence of signal cross sections. The leading order (LO) MSTW2008LO~\cite{MSTW,MSTW1} PDFs are chosen for all the signal samples. This PDF set provides a conservative estimate of the signal cross section at high BH masses with respect to the modern NNPDF3.0~\cite{NNPDF} PDFs. The cross section at large BH masses obtained with different PDF sets can vary up to an order of magnitude. The use of the MSTW2008LO PDF corresponds to the cross section at the lower edge of the uncertainty range for the NNPDF3.0 PDFs, therefore indicating a reasonable and conservative choice for signal simulation. The MSTW2008LO PDF has been recently superseded by the MMHT2014LO~\cite{MMHT} set. However, a numerical comparison of the signal cross sections computed using both of these PDF sets reveals no differences. This is expected, as no new data constraining PDFs for processes with such high momentum transfer have been utilized in the global fit used to extract the MMHT2014LO PDF set. Given that MSTW2008LO was the PDF of choice for the Run 1 version of this analysis, the use of this PDF also allows one to directly compare Run 1 and Run 2 results.

The CMS detector simulation is performed using both detailed simulation via {\GEANTfour}~\cite{Geant4} (semiclassical \BLACKMAX and \CHARYBDIS2 C1--C3 BH samples) and fast parametric simulation via the \textsc{FastSim}~\cite{FastSim} package (QBH and SB samples, as well as \CHARYBDIS2 C4--C5 samples). The fast simulation samples are validated against the full simulation, and the small observed differences between the two approaches were found to have a negligible impact on the signal acceptance.

In addition we use simulated samples of W+jets, Z+jets, $\gamma$+jets, $\ttbar$, and QCD multijet events for auxiliary studies. These events are generated with the  {\sc MadGraph5\_aMC@NLO} 2.2.2~\cite{MadGraph} event generator, followed by {\PYTHIA} for description of hadronization and fragmentation. The NNPDF3.0 PDF set with the tune CUETP8M1 is used for the background generation, and the CMS detector response is simulated via {\GEANTfour}.

\section{Backgrounds\label{s:backgrounds}}

\begin{figure}[htbp]
 \centering
  \includegraphics[width=0.49\textwidth]{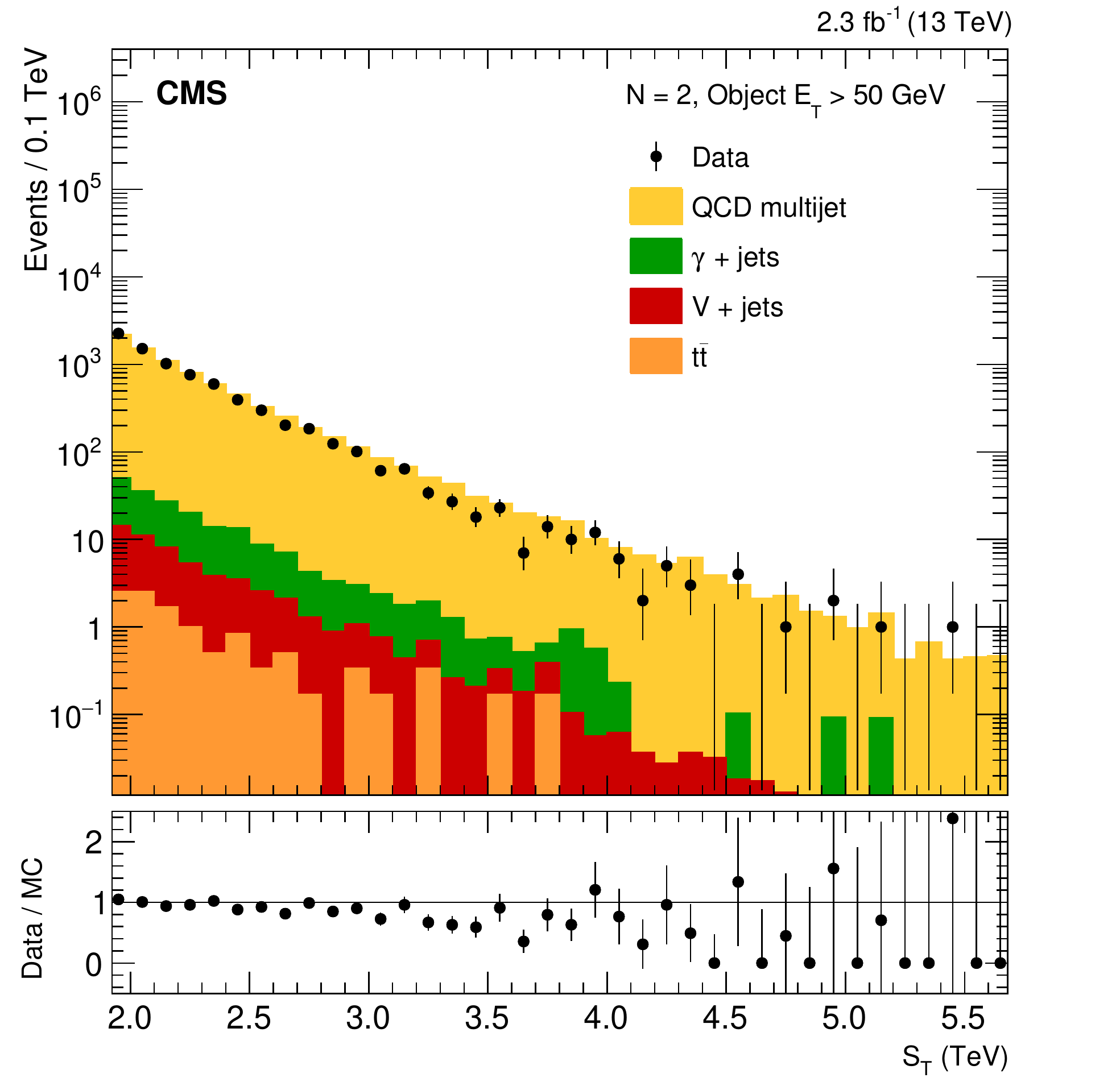}
  \includegraphics[width=0.49\textwidth]{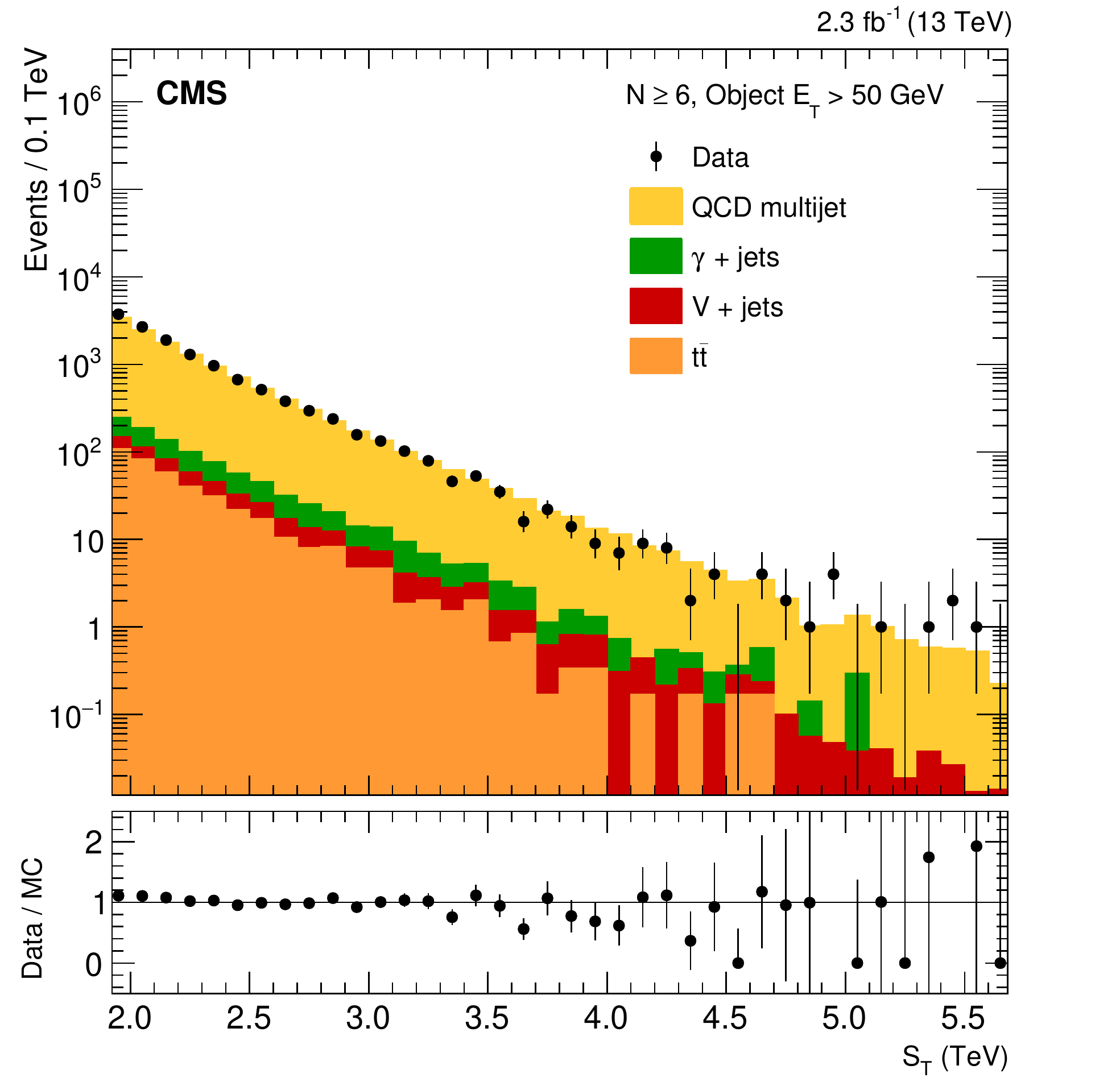}
  \caption{Contributions of the main QCD multijet background, as well as \photonjet, V+jets (V = \W,  \Z),  and ${\ttbar}$ backgrounds to the \st distribution for (\cmsLeft) exclusive multiplicity $N=2$ and (\cmsRight) inclusive multiplicity $N \ge 6$. All background predictions are based on simulated samples.}
   \label{fig:nonQCD}
 \end{figure}

The main SM backgrounds to the multiparticle final states from new-physics processes are QCD multijet and $\gamma$+jets production, vector bosons produced in association with jets (V+jets, where V stands for a W or Z boson), and ${\ttbar}$ process. The QCD multijet background is by far the dominant one. The additional backgrounds are not explicitly considered for the remainder of the analysis, as they together contribute only a few percent of the total background. They also have a very similar shape to the QCD multijet background, as evident from Fig.~\ref{fig:nonQCD}, which shows the comparison in the \st distribution of the total contribution of all the simulated backgrounds and the data for both the low-multiplicity $N=2$ and high-multiplicity $N \ge 6$ selections. While simulated samples are not used to estimate the background in this analysis, we note that the simulation nevertheless describes data reasonably well.

The background estimation strategy hinges on the shape of the \st distribution in the tail region of the spectrum being independent of the object multiplicity. As a consequence of this invariance, the \st distribution in higher multiplicity regions can be obtained by appropriately rescaling the low-multiplicity \st spectrum, where signal contamination is expected to be minimal. Given the similarity of the shapes of the QCD multijet and other backgrounds, this technique also implicitly accounts for most of the contribution of the non-QCD backgrounds.  The technique has been extensively validated in previous CMS publications~\cite{CMSBH1,CMSBH2,CMSBH3}, using both low-multiplicity data samples and simulated QCD multijet events.

The shape of the background is obtained by performing a binned likelihood fit of the \st spectrum to a functional form given by $P_{0}(1+x)^{P_{1}}x^{-P_{2} - P_{3}\log(x)}$, where the $P_i$ are free parameters of the fit, in the range between 1.4 and 2.4\TeV. The functional form used to fit the background is taken from earlier searches at the LHC and at the Fermilab Tevatron~\cite{PhysRevD.79.112002}, and represents a well-established characterization of rapidly  falling $\ET$ spectra. Since the goal is to operate in the background-only control region for the extraction of these shapes or templates, the multiplicities chosen for the background extraction are $N = 2$ and 3. This choice is justified by the dedicated Run 2 analyses (e.g.,~\cite{dijet-13}) as well as by the earlier iterations of this analysis (e.g.,~\cite{dijet-Run1}) where no signal of new physics was observed in these low-multiplicity regions.

The background template extracted by fitting the \st spectrum corresponding to the exclusive object multiplicity of $N = 2$ is normalized appropriately to obtain an estimate of the background for the \st spectra at higher multiplicities. The normalization regions in \st depend on the object multiplicity. Their definition is based on the studies of the \st invariance in simulated QCD multijet samples. The lower \st bound of the normalization region is chosen such that it is above the \st turn-on region, while the choice of the upper \st bound is guided by the need to operate in a regime of low signal contamination, namely where one does not expect a significant event yield for signals that have cross sections below those already excluded by the previous CMS search~\cite{CMSBH3}. The normalization factors are calculated as the ratio of the number of data events in the normalization regions for the inclusive multiplicities of $N\geq 2\dots 10$ and that for the exclusive multiplicity of $N = 2$. These factors, along with the choice of the normalization region, are detailed in Table~\ref{tab:NF}.

\begin{figure*}[htbp]
\centering
\includegraphics[width=0.45\textwidth]{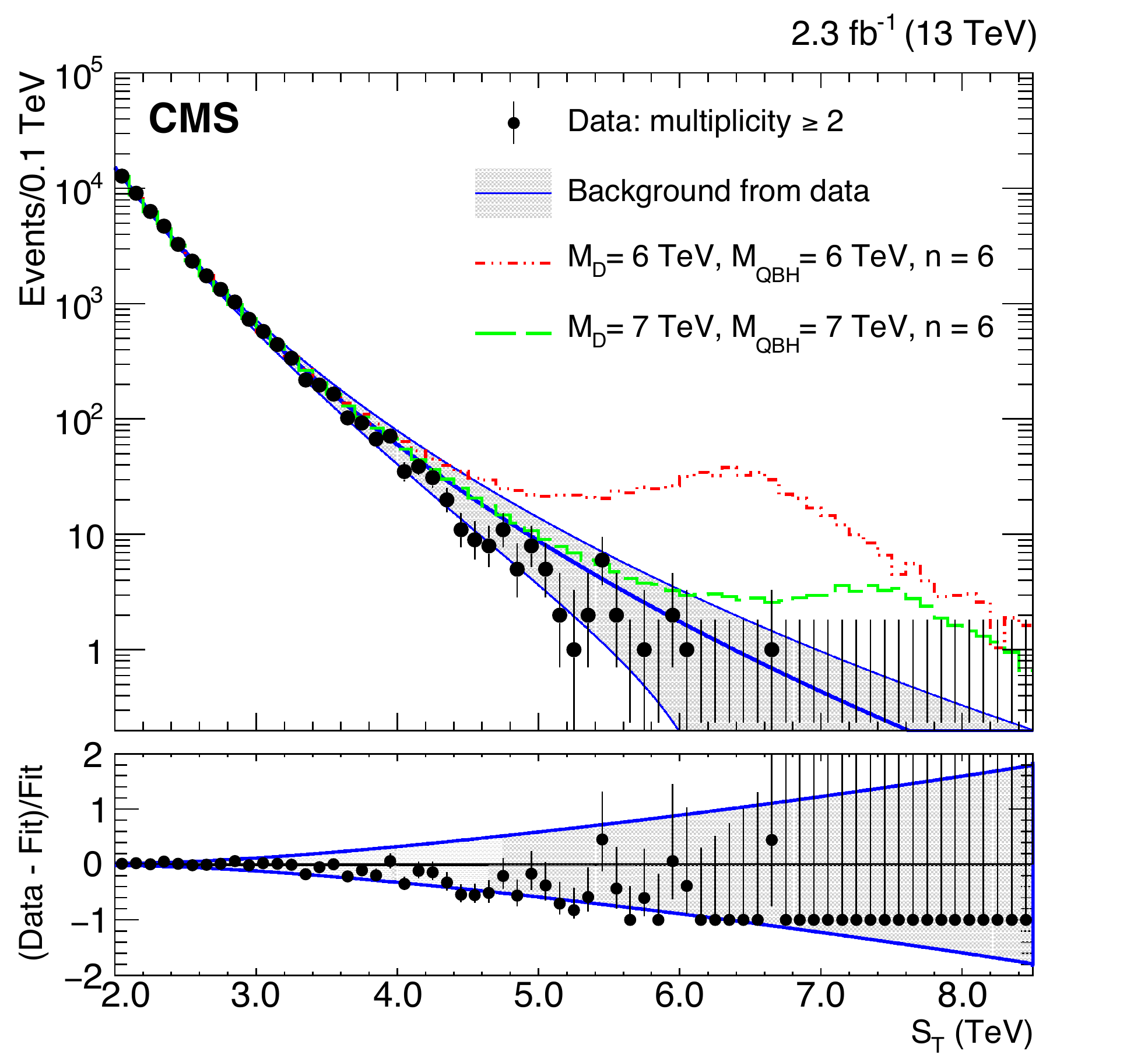}
\includegraphics[width=0.45\textwidth]{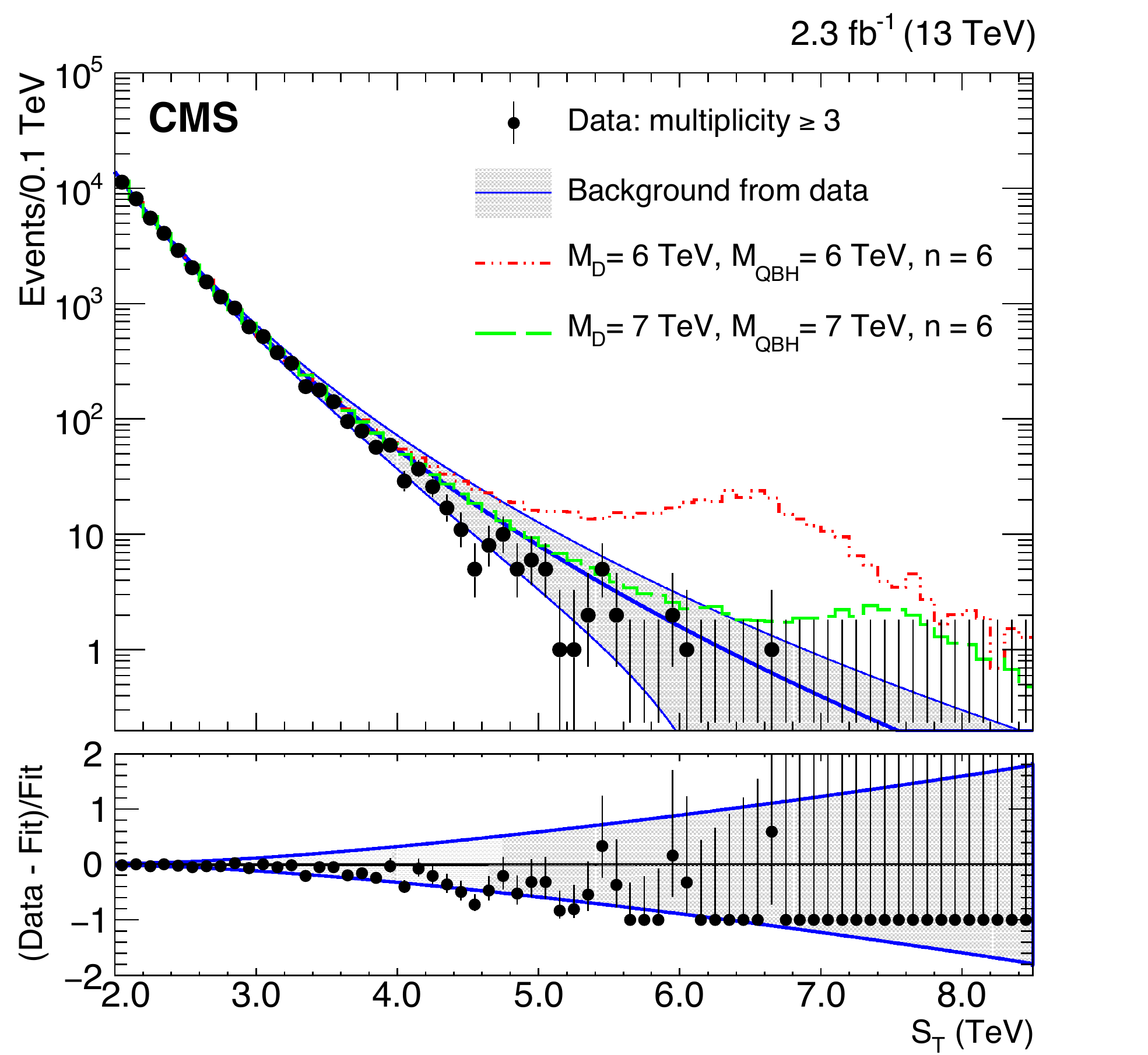}
\includegraphics[width=0.45\textwidth]{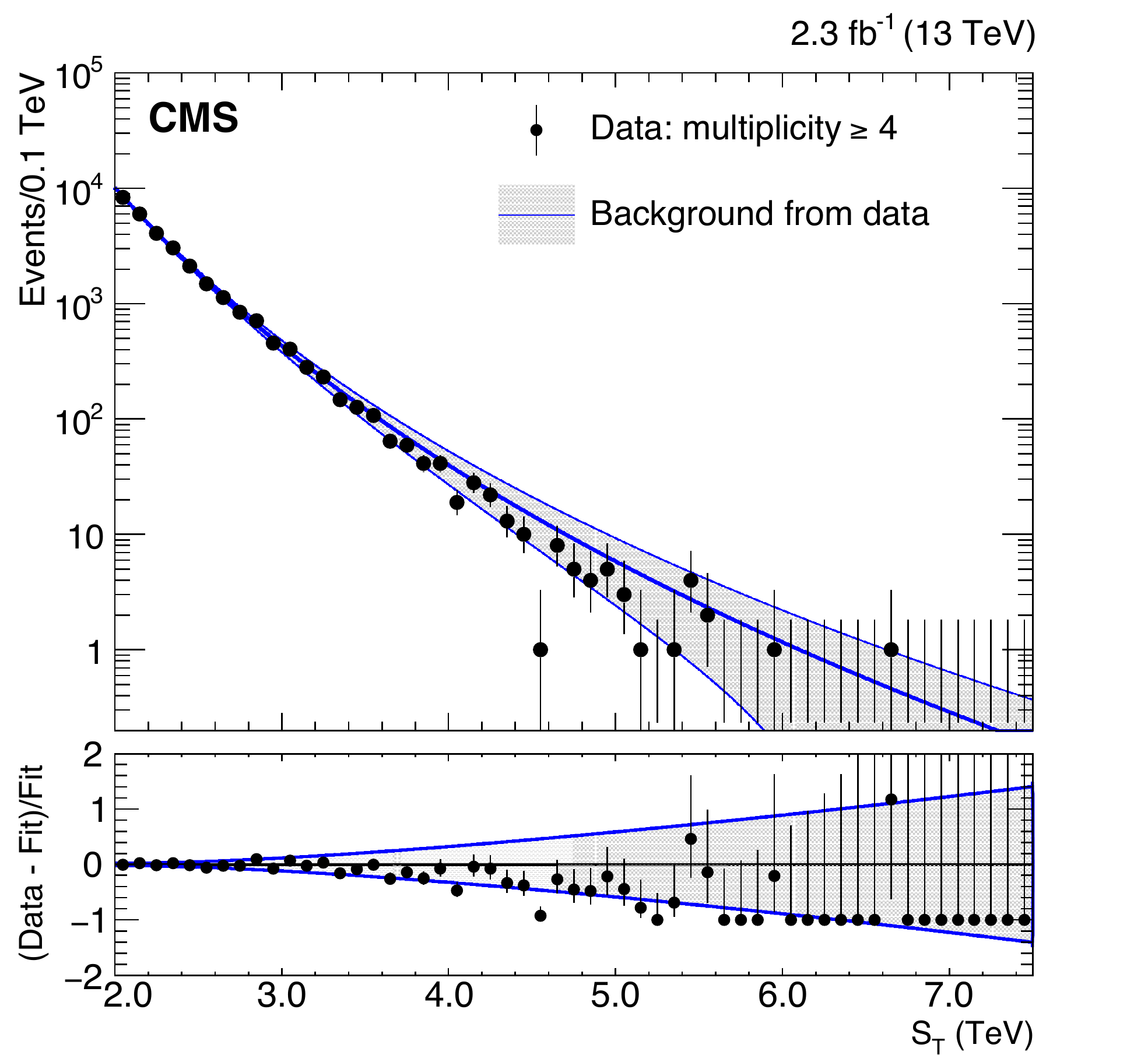}
\includegraphics[width=0.45\textwidth]{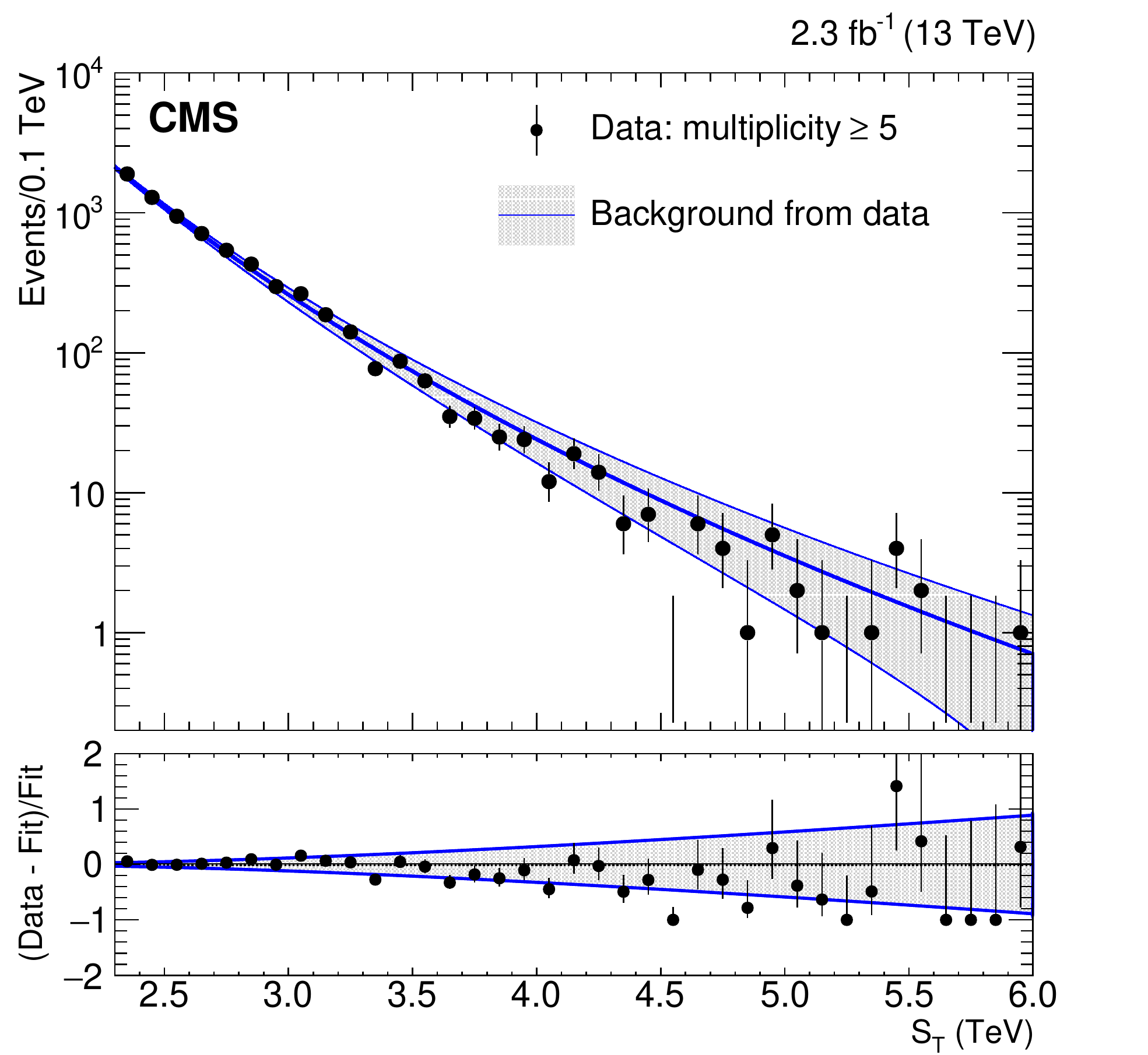}
\caption{The distributions of the total transverse energy, \st, for inclusive multiplicities of objects (electrons, muons, photons, or jets) $N\geq 2, 3, 4, 5$. Observed data are shown by points with error bars, the solid blue lines along with the gray shaded band show the main background estimation (central blue line), along with the uncertainty band (outer blue lines). The deviation of the fit from the data is shown in each lower pane. The top two
plots also show predictions for two quantum black hole benchmark scenarios added to the corresponding background predictions.}
\label{fig:stn3n6}
\end{figure*}

\begin{figure*}[htbp]
\centering
\includegraphics[width=0.45\textwidth]{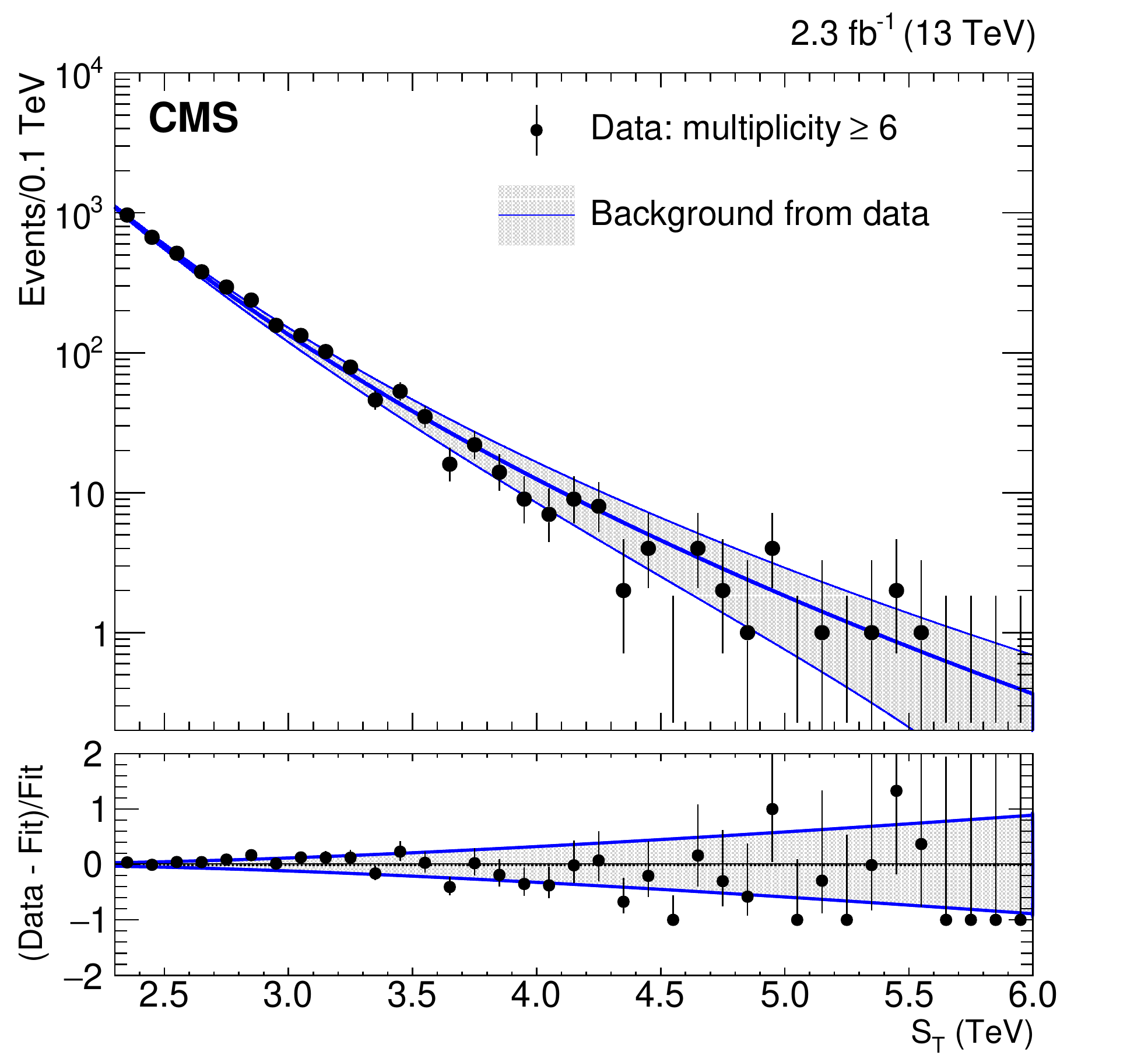}
\includegraphics[width=0.45\textwidth]{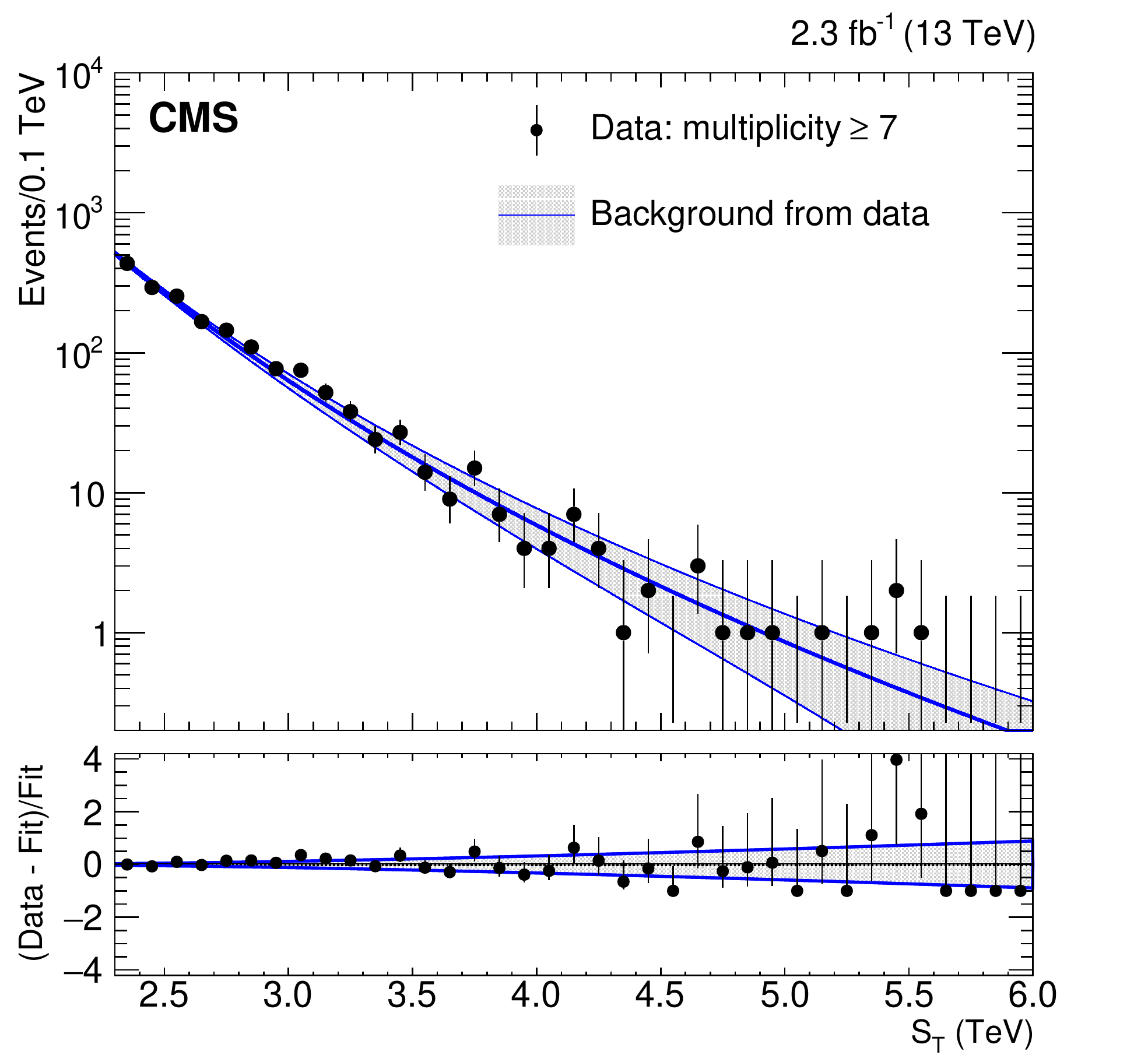}
\includegraphics[width=0.45\textwidth]{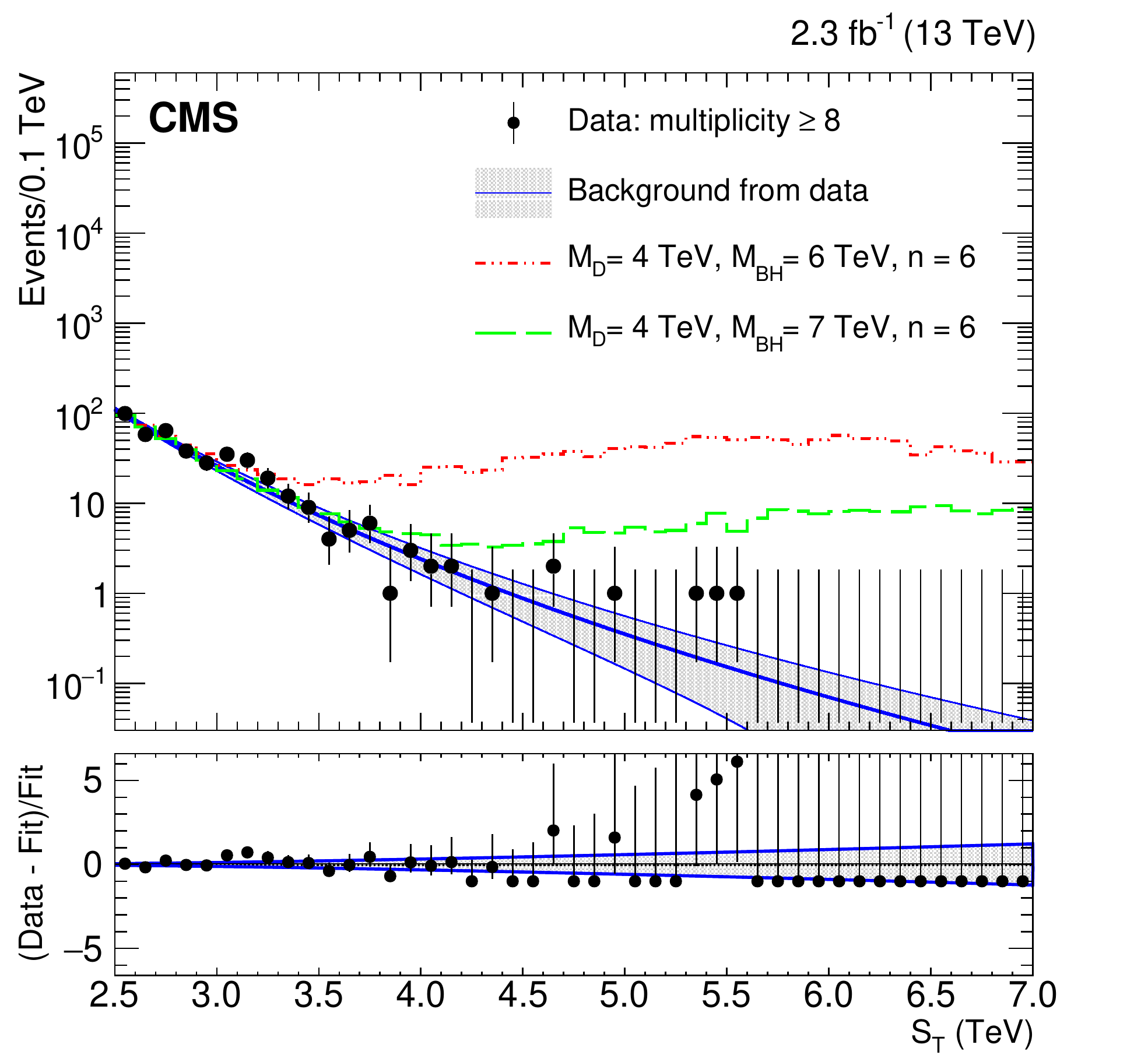}
\includegraphics[width=0.45\textwidth]{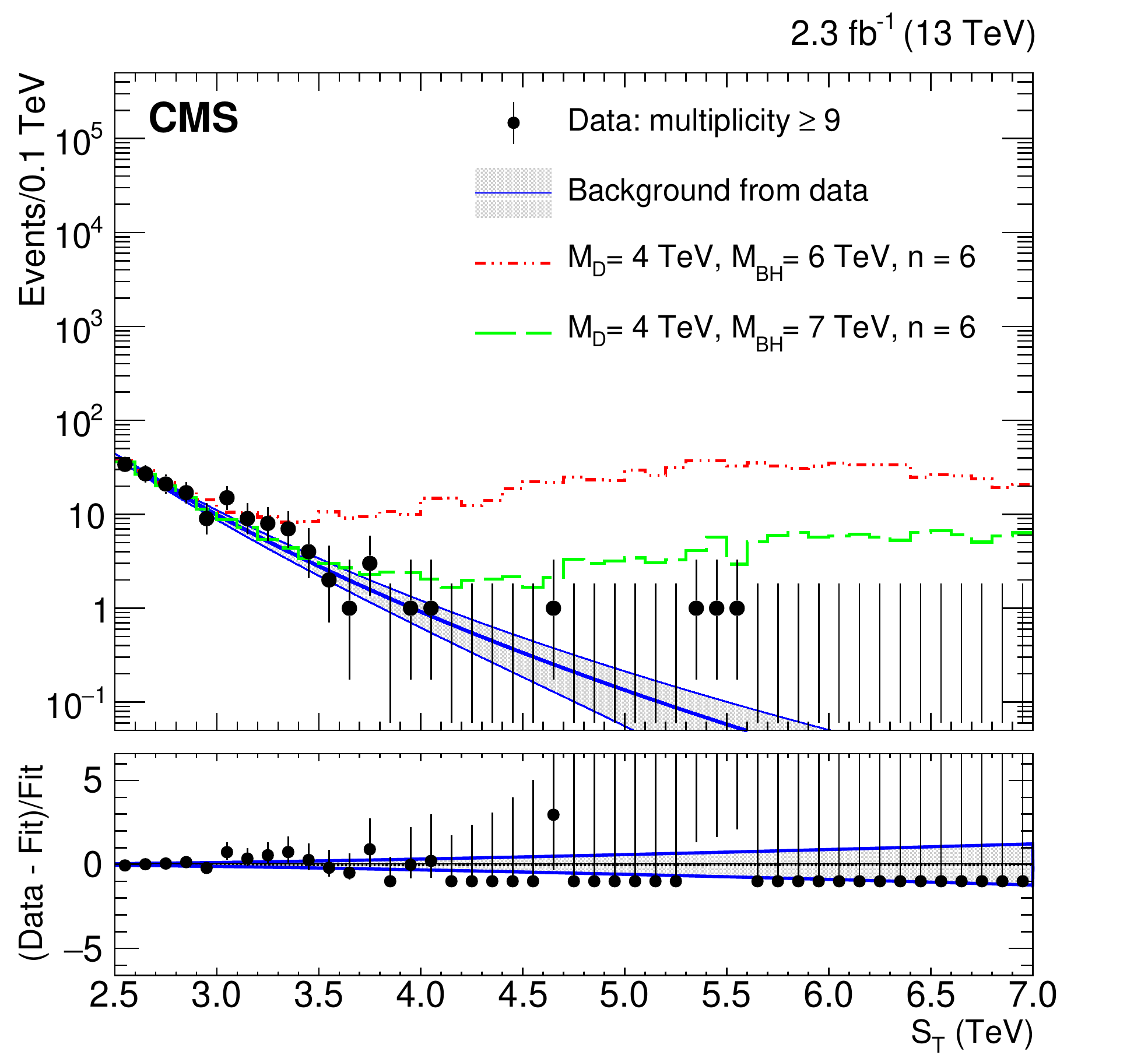}
\includegraphics[width=0.45\textwidth]{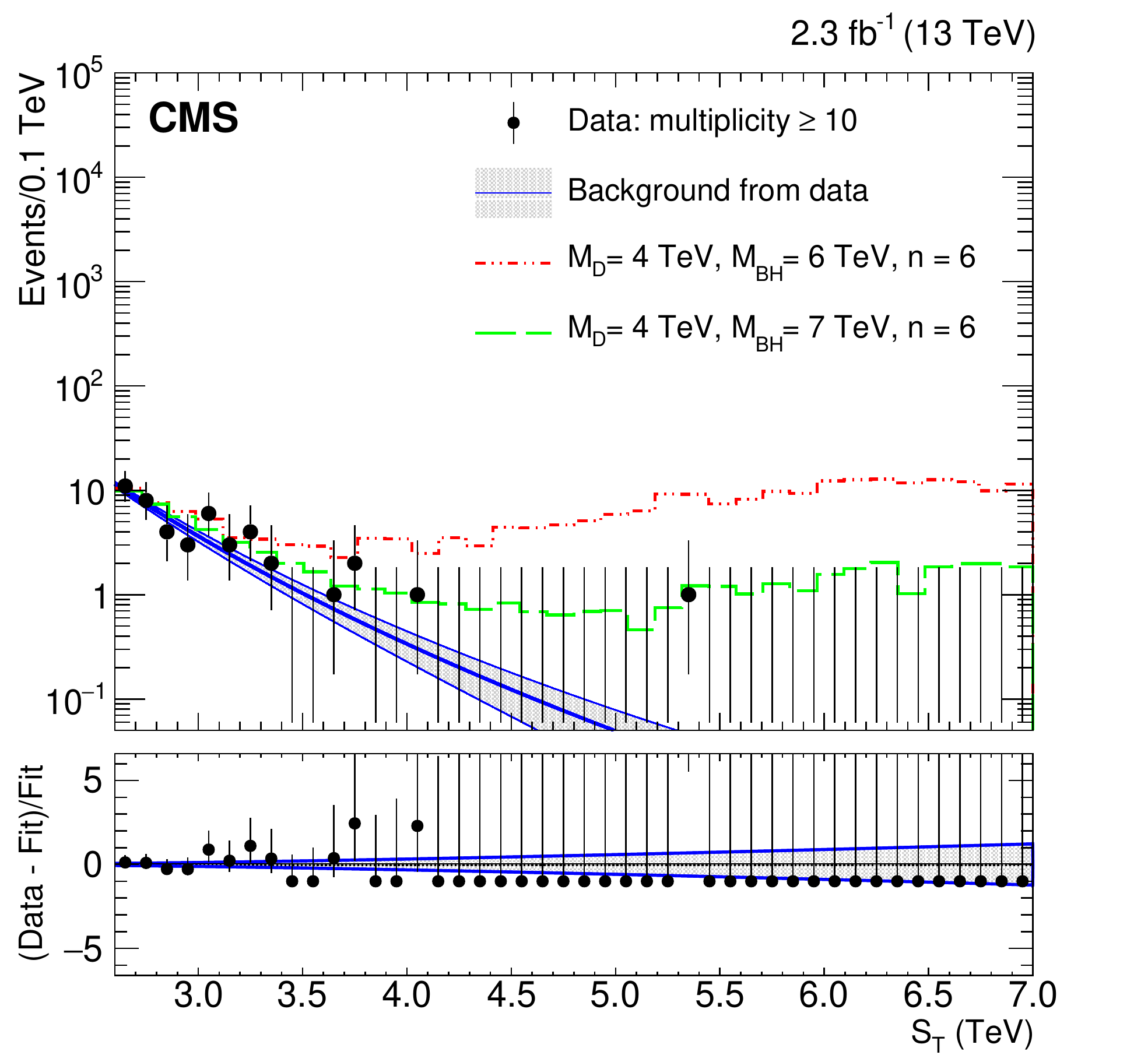}
\caption{The distributions of the total transverse energy, \st, for inclusive multiplicities of objects (photons, muons, photons, or jets) $N\geq 6, 7, 8, 9, 10$. Observed data are shown by points with error bars, the solid blue lines along with the gray shaded band show the main background estimation (central blue line), along with the uncertainty band (outer blue lines). The deviation of the fit from the data is shown in each lower pane. The lower three plots also show predictions for two semiclassical black hole signal benchmarks added to the corresponding background predictions. }
\label{fig:stn7n10}
\end{figure*}

To ascertain the uncertainties associated with this method of background extraction, two additional fitting functions are considered, given by $P_{0}(P_{1}+x)^{-P_{2}}$ and $P_{0}(P_{1}+P_{2}x+x^{2})^{-P_{3}}$. With an aim to compute a conservative estimate of the uncertainty, both the $N =$ 2 and 3 \st spectra are used to estimate the background shape. This leads to six different functions used in the fit: one nominal and five additional ones that form an uncertainty envelope that is symmetrized around the main fit. The shape systematic uncertainty is indicated with the gray shaded bands in Figs. \ref{fig:stn3n6} and \ref{fig:stn7n10}, which show the \st spectrum observed in data for various inclusive multiplicities and the background predictions with their uncertainties. It ranges between 1 and 200\% and rapidly increases in the high-\st range because of the limited number of events in the $N =$ 2 and 3 \st spectra used to derive the background templates. We assign an additional 5\% systematic uncertainty to address possible deviation from the \st invariance by estimating the difference between the background predictions based on the $N = 2$ (default) and $N = 3$ templates. This is a subdominant uncertainty in the high-\st range relevant for the analysis. The systematic uncertainties in the background estimate are detailed in Table~\ref{tab:Table_Uncertainties} of Section~\ref{sec:systematicUncertainties}.

As seen from Figs.~\ref{fig:stn3n6} and \ref{fig:stn7n10}, the data agree well with the predicted background and no evidence for new physics production  is observed in any of the multiparticle final states studied.

\begin{table}[htbp]
\centering
\topcaption{The normalization regions and the corresponding $N = 2$ background template normalization factors $s$ and their uncertainties for inclusive multiplicities, $N\geq 2\ldots 10$. The normalization factor uncertainties are given by $s/\sqrt{N_\mathrm{NR}}$, where $N_\mathrm{NR}$ is the number of events in each normalization region.}
\begin{tabular}{ccc}
Multiplicity & \multicolumn{2}{c}{Normalization}  \\\cline{2-3}
& Region [\TeVns{}] & Factor \\
\hline
 $\ge$2 & 2.0--2.3  &  8.66 $\pm$ 0.15 \\
 $\ge$3 & 2.0--2.3  &  7.66 $\pm$ 0.13 \\
 $\ge$4 & 2.0--2.3  &  5.67 $\pm$ 0.10 \\
 $\ge$5 & 2.3--2.6  &  3.28 $\pm$ 0.09 \\
 $\ge$6 & 2.3--2.6  &  1.71 $\pm$ 0.05 \\
 $\ge$7 & 2.3--2.6  &  0.770 $\pm$ 0.022\\
 $\ge$8 & 2.5--2.8  &  0.330 $\pm$ 0.013 \\
 $\ge$9 & 2.5--2.8  &  0.124 $\pm$ 0.005 \\
 $\ge$10 & 2.6--2.9  &  0.047 $\pm$ 0.002\\
 \end{tabular}
\label{tab:NF}
\end{table}
s

\section{Systematic uncertainties\label{sec:systematicUncertainties}}

A detailed study of the systematic uncertainties associated with this analysis was carried out, and these uncertainties were taken into account while computing exclusion limits, as detailed in Section~\ref{sec:limits}. The dominant sources of systematic uncertainties arise from the effects of the jet energy scale (JES) and JER uncertainties, from the choice of the PDFs, and from the migration between event categories as a result of final-state radiation (FSR). The uncertainties vary for different signal samples and different inclusive multiplicities. The following sections give details on their estimation and the typical range for each uncertainty.

\subsection{Parton distribution functions}

There are two different sources of the PDF uncertainty that could affect the signal acceptance. The uncertainty could simply come from the choice of the PDF set.  Alternatively, the systematic uncertainty could come from the variations induced by the uncertainties associated with the fit parameters in a chosen PDF set. The latter uncertainty is estimated by calculating $2k+1$ weights per event, where $k$ is the number of alternative PDF sets with parameters varied within their associated uncertainties either up or down. These alternative sets are provided by the authors of each PDF set, along with the main PDFs used in the signal generation. The signal acceptance is then calculated for each of these weights, resulting in $2k+1$ values of the acceptance. For a particular choice of PDF set, the systematic uncertainty in the acceptance is quoted as the sum in quadrature of the deviations from the central value for each of the $k$ PDF eigenvectors. This analysis is further extended by using various PDFs, such as MSTW2008LO, CTEQ6.1L~\cite{CTEQ6L}, and CT10~\cite{CT10}. The uncertainty is computed for a particular benchmark scenario, a nonrotating BH with \MD = 3\TeV, \mbh = 5.5\TeV, and $n = 2$, representative of the uncertainties for other benchmark points in the range probed by this analysis. The uncertainty in the acceptance using the variation of the chosen PDF, MSTW2008LO, does not exceed 0.5\%. The uncertainty due to the choice of the PDF is significantly larger and can be as high as 6\%. Following the recommendations of the PDF4LHC group~\cite{pdf4lhc,pdf4lhc1}, we assign a total uncertainty of 6\% associated with this source of the systematic uncertainty.

\subsection{Jet energy scale and resolution}

The CMS experiment has adopted a factorized approach~\cite{Chatrchyan:2011ds,Khachatryan:2016kdb} to the application of corrections associated with the energy of a jet. After the subtraction of the additional energy due to pileup, the energies of the reconstructed jets are corrected for the nonlinear response of the calorimeter. These corrections are parametrized in \pt and $\eta$, and are derived from simulation and in situ measurements of the energy balance in dijet, mulitjet, $\gamma$+jet, and leptonic $\PZ$+jet events~\cite{Chatrchyan:2011ds,Khachatryan:2016kdb}. Given the predominance of jets in the final state of interest, the uncertainty related to the JES can significantly affect the signal acceptance. The JES is modified by one standard deviation in both upward and downward directions, and the corresponding changes in the jet energies are propagated into the \MET and \st computation. For various model-specific scenarios, we find the variation to lie between 1 and 5\% and conservatively assign a uniform uncertainty in signal acceptance of 5\% due to the JES uncertainty.

The uncertainty due to the JER is estimated by oversmearing jet energies in simulated signal samples with $\eta$-dependent factors to match the resolution observed in data. The effect of this change to the jet energy is propagated to the \MET and \st calculations. The JER uncertainty in the signal acceptance varies between 1.2 and 4.0\%; we therefore assign a conservative uniform uncertainty of 4\% to account for this effect.

In estimating both the JES and JER uncertainties, we find that there are migrations of events in different multiplicity categories due to the jet \pt values moving either above or below the 50\GeV threshold. The respective values of uncertainties associated with both of these sources sufficiently cover the effect of event migration. These two uncertainties affect only the simulated signal yields, as the background determination does not rely on simulation.

\subsection{Final-state radiation}

The presence of imperfectly modeled FSR may result in event migration between various multiplicity bins due to the change in the number of objects counted toward the \st. The uncertainty in the signal acceptance due to the imperfect modeling of FSR is estimated by varying the parameters in the {\PYTHIA 8} generator that govern FSR modeling, and is found to be 1.2\%.

The sources of systematic uncertainties considered in the analysis and their magnitude are listed in Table~\ref{tab:Table_Uncertainties}.

\begin{table*}[htbp]
  \centering
    \topcaption{Summary of the systematic uncertainties. The range of the background uncertainties correspond to the \st range probed. A dash implies that the corresponding uncertainty source does not apply.}
    \label{tab:Table_Uncertainties}
    \begin{tabular}{lcc}
        Uncertainty             & Effect on signal acceptance   & Effect on background estimate \\ \hline
        JES     & $\pm$5\%                     & \NA \\
        JER     & $\pm$4\%                     & \NA \\
        PDF                     & $\pm$6\%                     & \NA \\
        FSR                  & $\pm$1.2\%              & \NA \\
        Integrated luminosity & $\pm$2.7\% & \NA \\
        Background normalization                & \NA                            & $\pm$(0.5--5.2)\%
        \\
        Background shape                & \NA                            & $\pm$(1--200)\%, \\
        Potential \st noninvariance & \NA & $\pm$5\%\\
    \end{tabular}
\end{table*}

In addition, an uncertainty in the integrated luminosity of 2.7\%~\cite{lumi} is propagated to both the model-independent and model-specific limits. All other uncertainties are negligible~\cite{CMSBH1,CMSBH2,CMSBH3}.
\section{Model-independent limits\label{sec:limits}}

No statistically significant deviations of data relative to the background predictions are observed in any of the spectra. Exclusion limits on various signals are set using the modified frequentist CL$_\mathrm{s}$ approach~\cite{Junk,Read} with the profile likelihood as a test statistic and nuisance parameters implemented via log-normal priors. The likelihood of the observed number of events is modeled with a Poisson distribution. The limit calculation is performed using the methodology developed by the ATLAS and CMS Collaborations in the search for the Higgs boson~\cite{HiggsCombination}. The systematic uncertainties taken into account are detailed in Section~\ref{sec:systematicUncertainties}.

\begin{figure*}[htbp]
   \centering
 \includegraphics[width=0.45\textwidth]{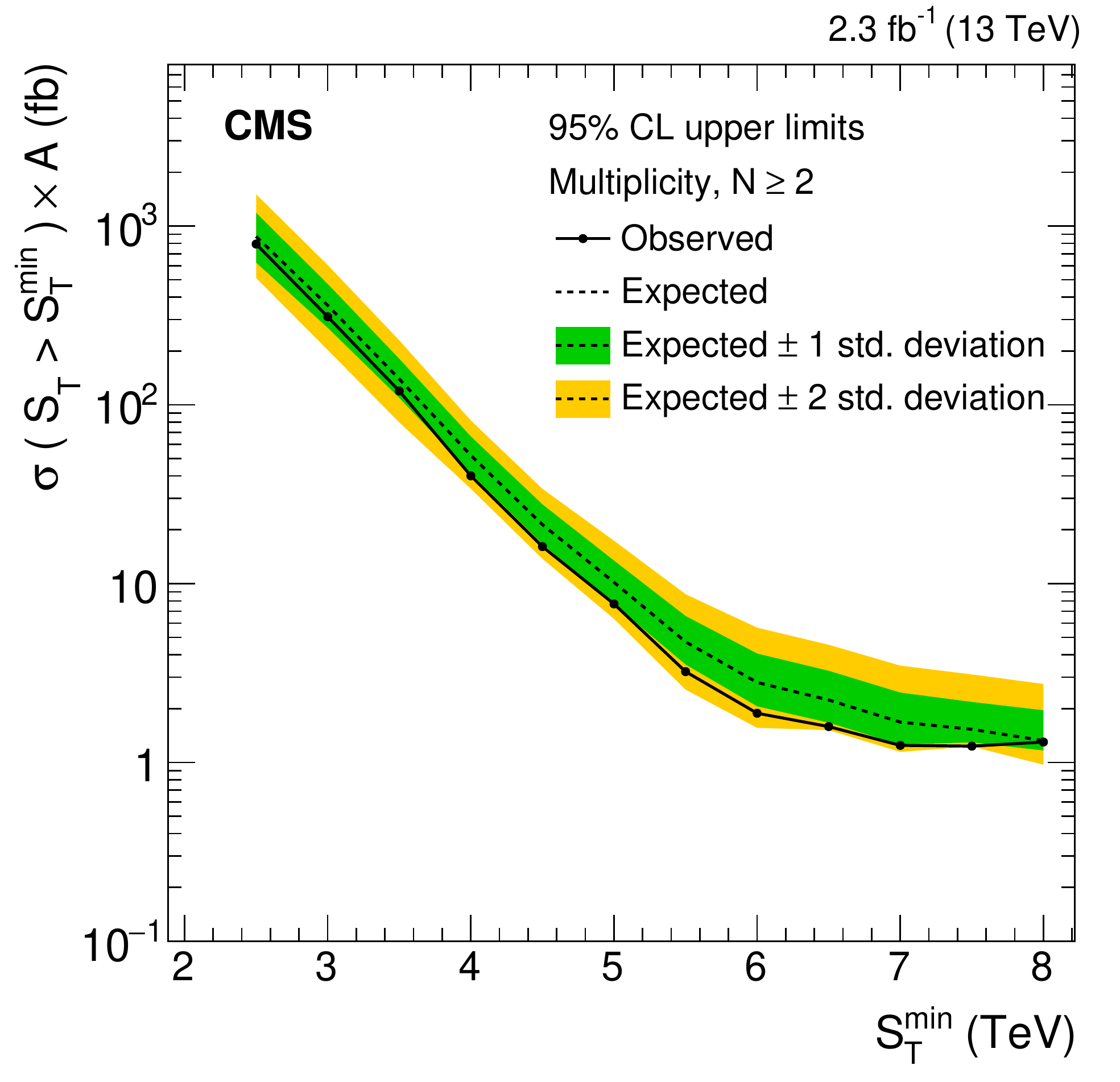}
 \includegraphics[width=0.45\textwidth]{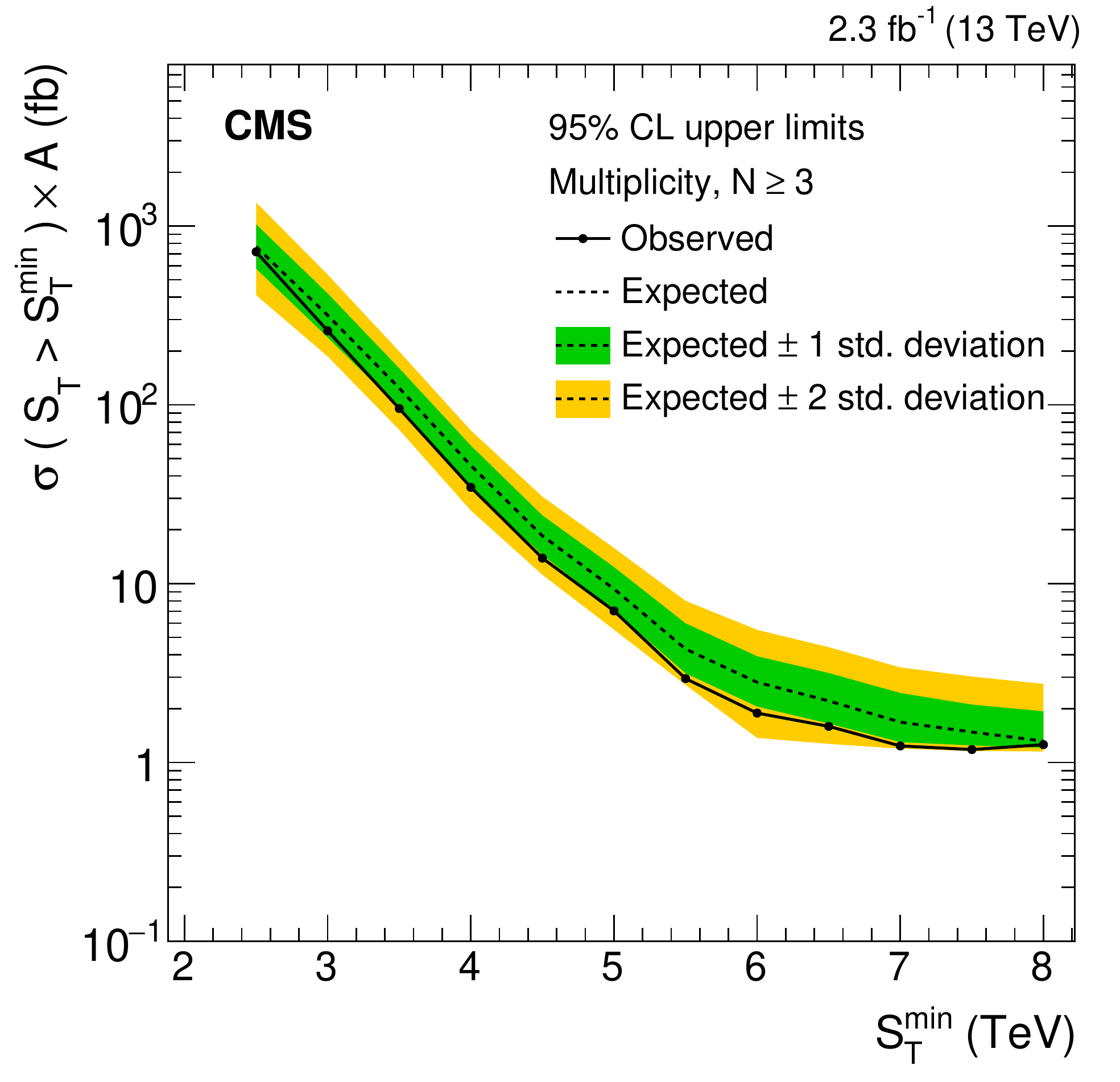}
 \includegraphics[width=0.45\textwidth]{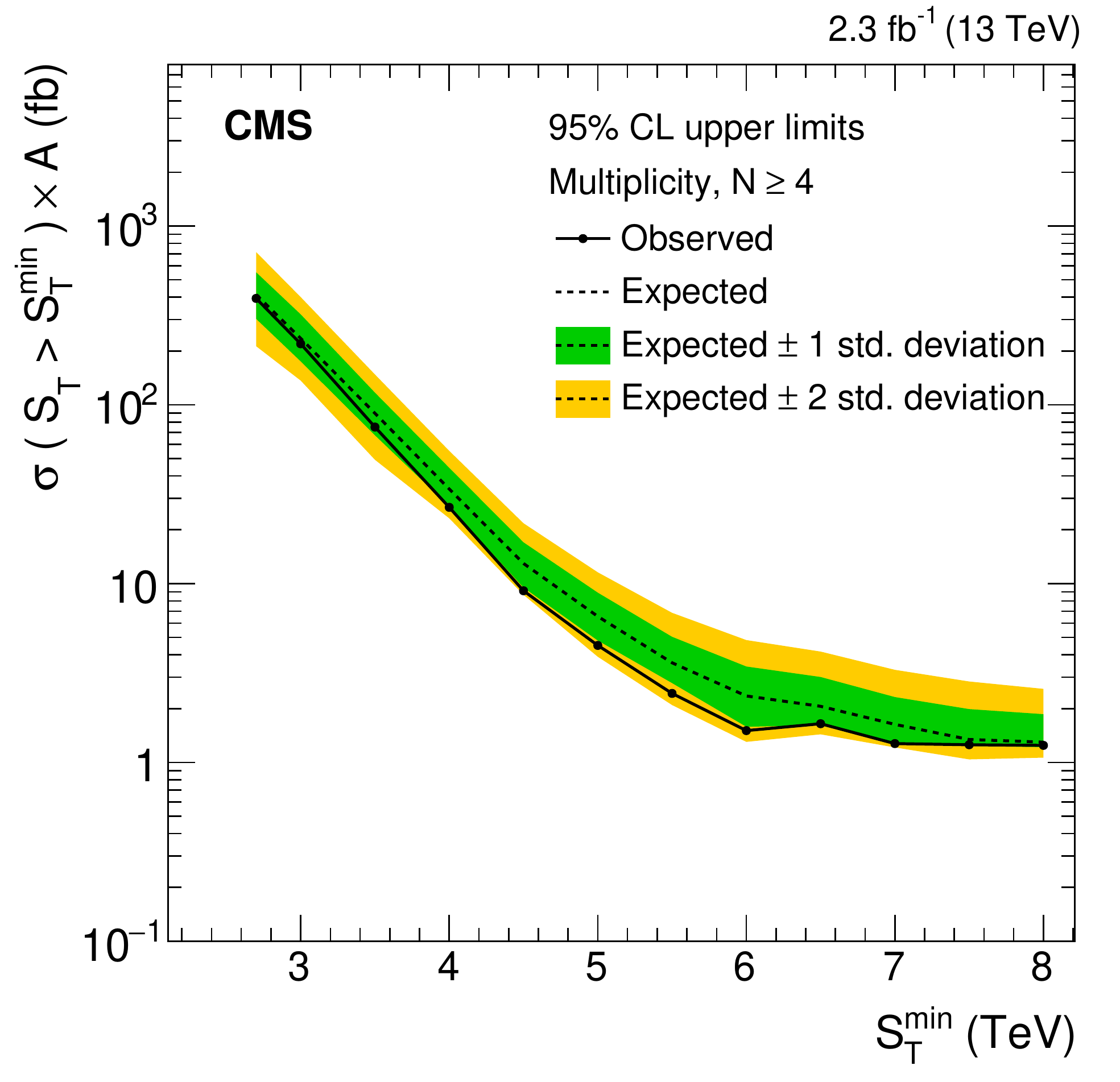}
 \includegraphics[width=0.45\textwidth]{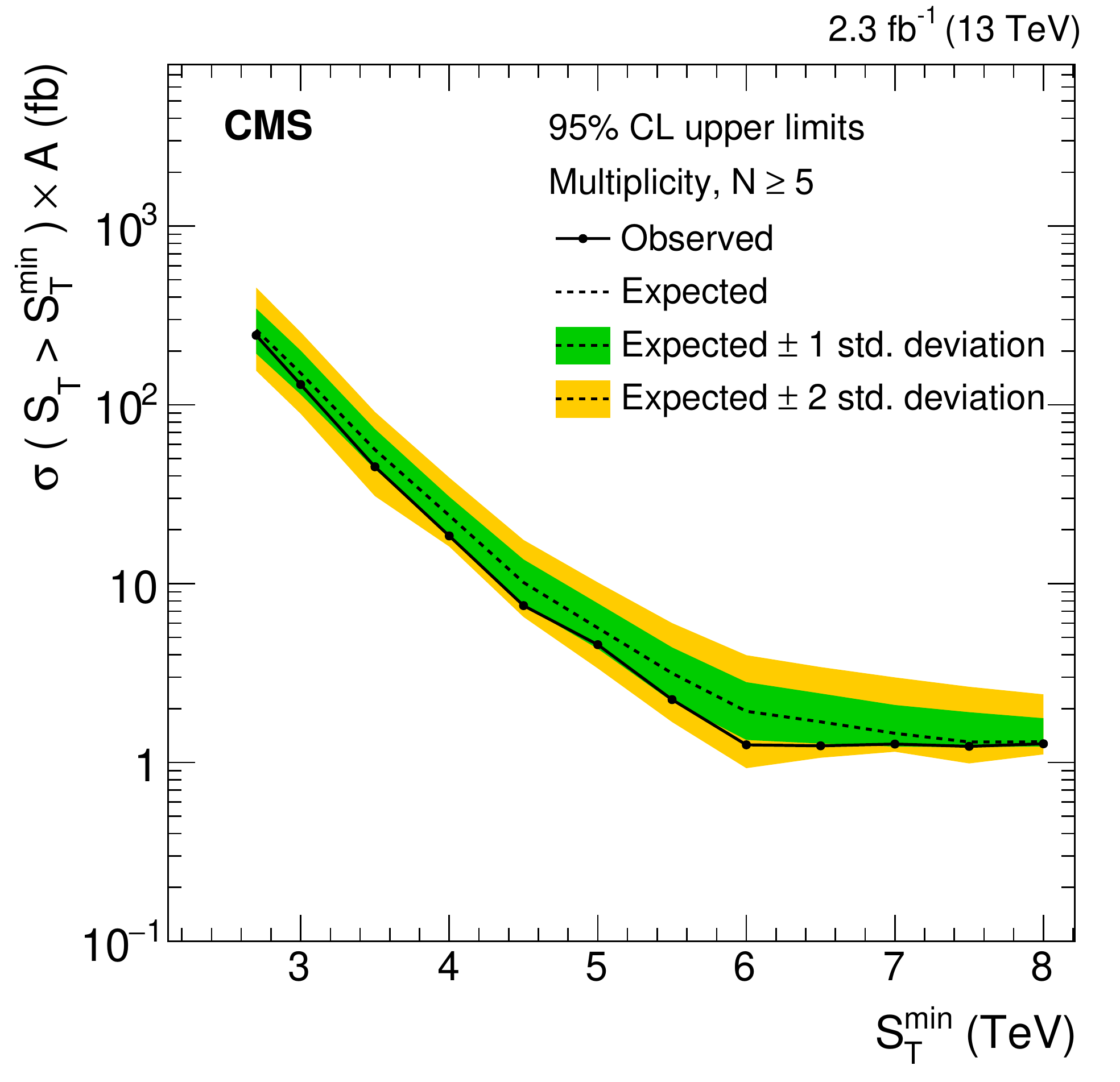}
   \caption{Model-independent 95\% CL upper limits on the cross section times acceptance for four sets of inclusive multiplicity thresholds: $N \ge 2$, 3, 4, and 5. Observed (expected) limits are shown as solid (dashed) lines, and the two bands correspond to $\pm$1 and 2 standard deviations in the expected limit.}
   \label{fig:Limit_ModelIndependent1}
\end{figure*}

\begin{figure*}[htbp]
   \centering
 \includegraphics[width=0.45\textwidth]{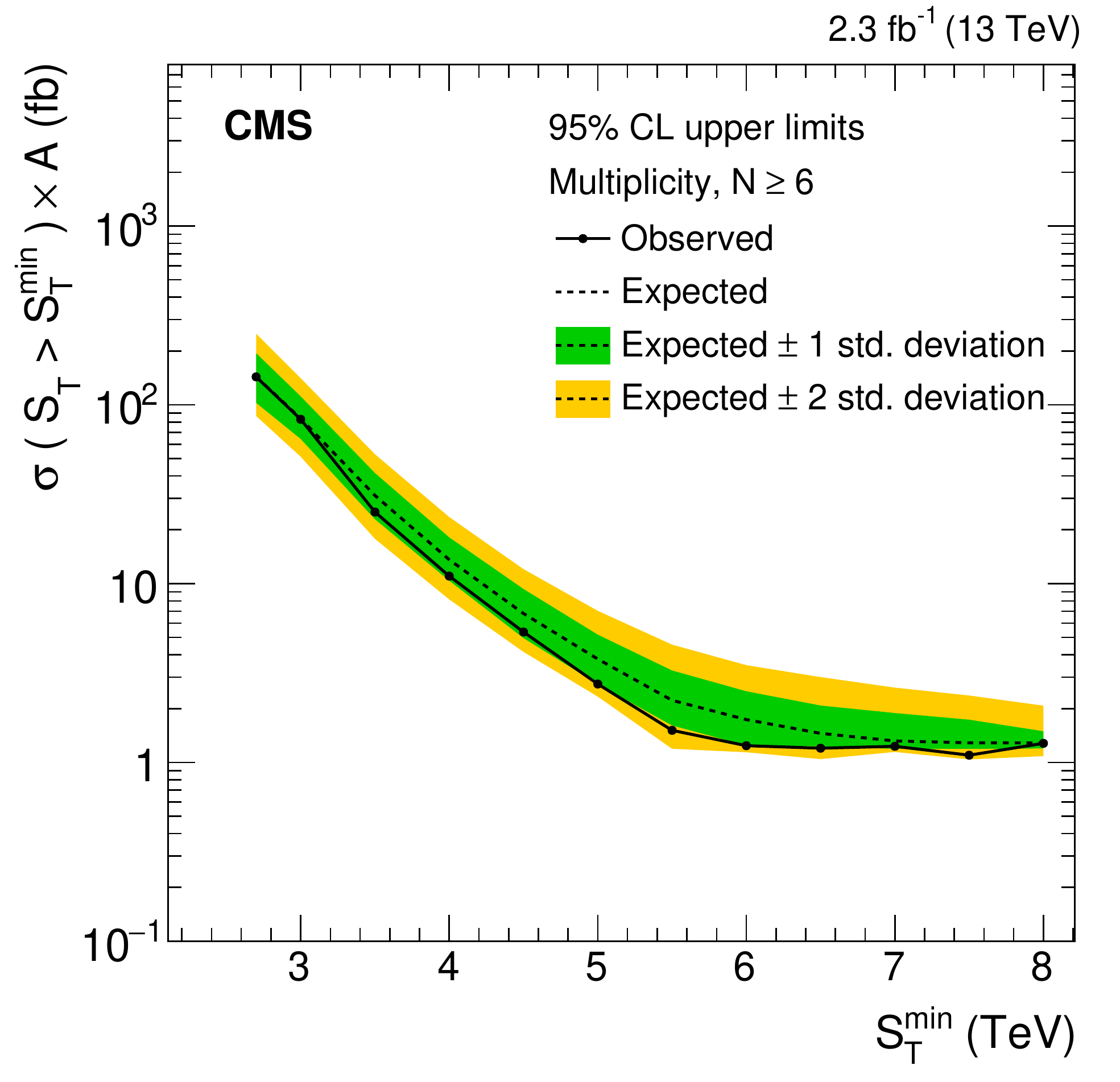}
 \includegraphics[width=0.45\textwidth]{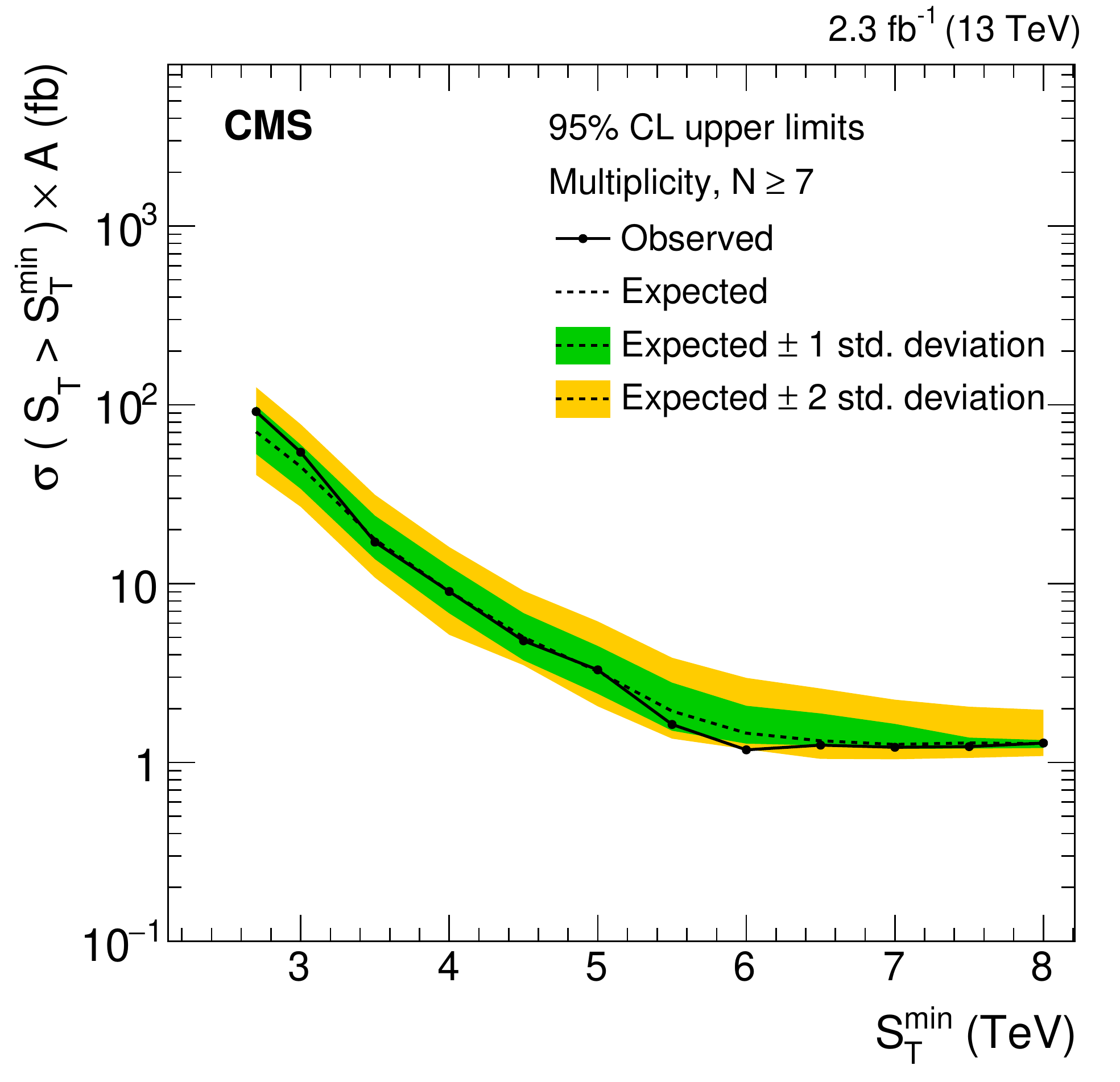}
 \includegraphics[width=0.45\textwidth]{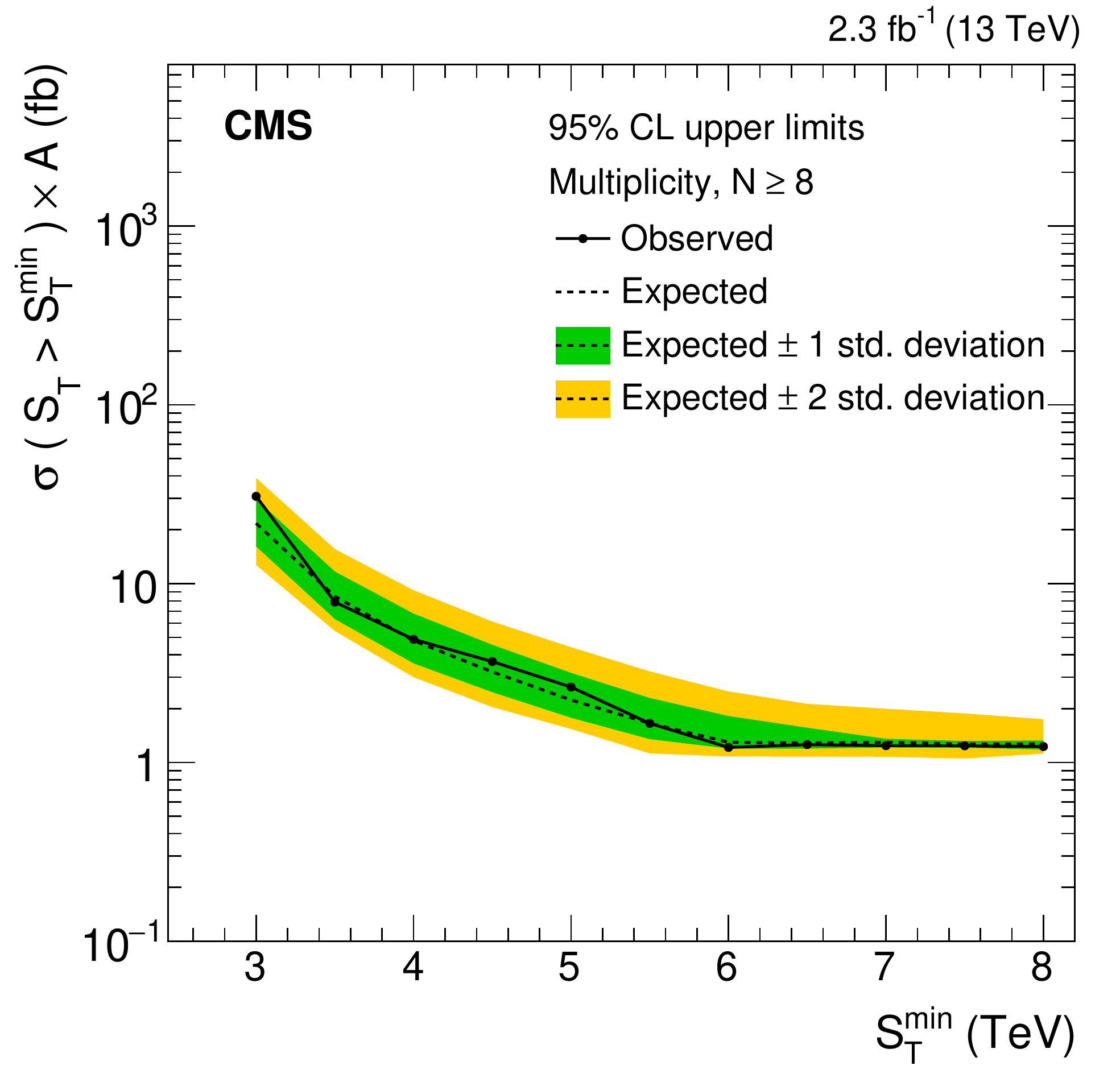}
 \includegraphics[width=0.45\textwidth]{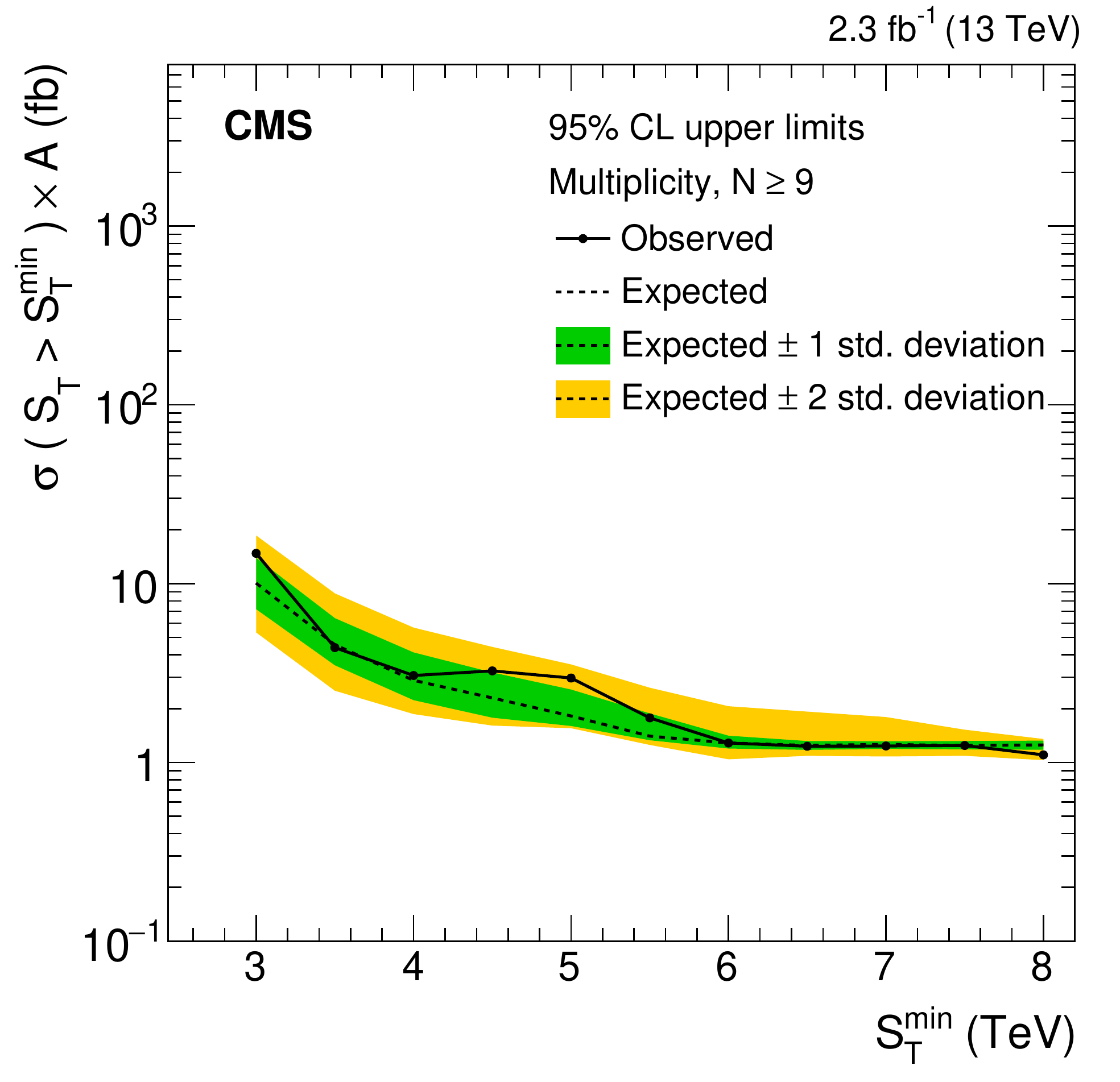}
 \includegraphics[width=0.45\textwidth]{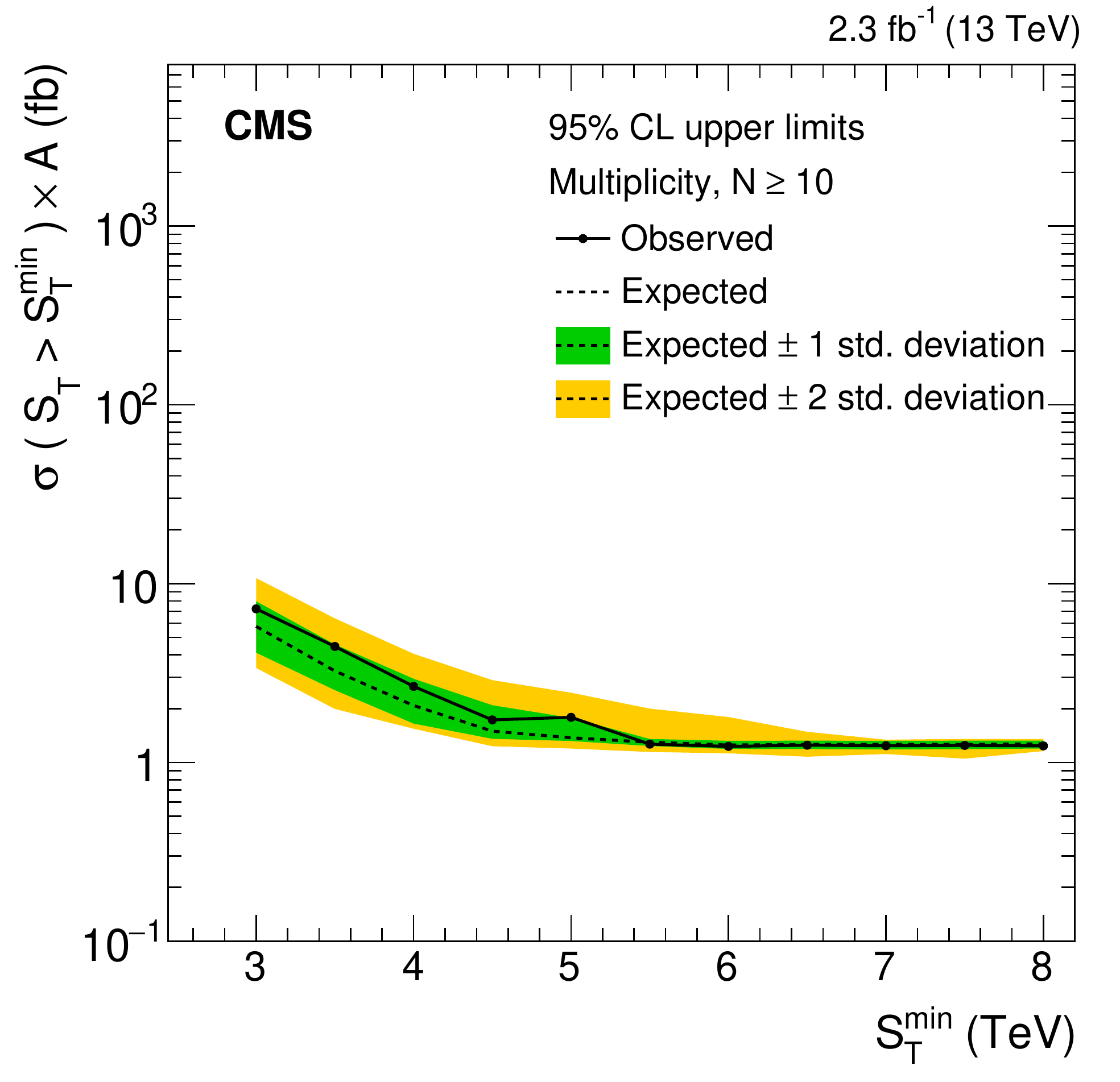}
    \caption{Model-independent 95\% CL upper limits on the cross section times acceptance for five sets of inclusive multiplicity thresholds: $N \ge 6$, 7, 8, 9, and 10. Observed (expected) limits are shown as a solid (dashed) lines, and the two bands correspond to $\pm$1 and 2 standard deviations in the expected limit.}
   \label{fig:Limit_ModelIndependent2}
\end{figure*}

As mentioned in Section~\ref{sec:introduction}, the main emphasis of the analysis is the computation of model-independent limits on the product of the hypothetical signal cross section and acceptance ($\sigma \, A$) in inclusive $N \ge N^\text{min}$ multiplicity bins, as a function of the minimum \st requirement ($S_\mathrm{T}^\text{min}$). These limits represent the minimum $\sigma \, A$ of a potential signal excluded by the analysis and do not assume any particular signal model. This makes the limits applicable to a large variety of BH models and other theoretical models with processes that result in high-multiplicity final states. To obtain the model-independent limits, a counting experiment is performed for $N \ge N^\text{min}$ and $S_\mathrm{T} > S_\mathrm{T}^\text{min}$. These limits, at 95\% confidence level (CL), are presented in Figs.~\ref{fig:Limit_ModelIndependent1} and \ref{fig:Limit_ModelIndependent2} for $N^\text{min} \ge 2,\ldots,10$. The limits on $\sigma \, A$ approach 1.3\unit{fb} at high values of $S_\mathrm{T}^\text{min}$.

Model-independent limits can be used straightforwardly to test any model of new physics that predicts signal corresponding to high-multiplicity, energetic final states. Given that the object selection efficiency is close to 100\%, all is needed to test a particular model against the results of this search is to estimate the signal acceptance as a function of the minimum \st requirement, corresponding to the selections described in Section~\ref{sec:event_reconstruction}, which can be done with sufficient precision even at the generator level. A comparison of the product of the signal cross section and the acceptance as a function of the minimum \st requirement with the model-independent limit at the relevant object multiplicity would then indicate whether a particular model has been excluded by this analysis.

\section{Model-specific limits}

In the case of the model-specific limits, the selection criteria ($N^\text{min}$, $S_\mathrm{T}^\text{min}$) were chosen for optimum sensitivity to each particular model. This was done by using the lowest value of the expected limit, as well as the maximum value of the Z$_\mathrm{bi}$~\cite{Zbi} statistic, and converting the corresponding model-independent limit into a limit on the BH or SB mass, using the known signal cross section and acceptance. The Z$_\mathrm{bi}$ statistic is a measure of equivalent Gaussian signal significance obtained by considering the binomial probability of the events in data being distributed at least as signal-like as observed, under the assumption of the background-only hypothesis. In the majority of cases, the limit- and significance-based optimizations are in agreement and lead to the same set of optimum selection requirements. Since the signal MC benchmark points were produced on a grid of discrete \mbh values (with a typical spacing of 1\TeV), in order to obtain the mass limit, we smoothly interpolate the signal cross section times acceptance and the exclusion limit between the adjacent points on the grid. Typical precision of this procedure expressed as a limit on the minimum BH mass is $\sim$0.1\TeV, although for some of the points where the grid spacing was not as fine, the precision can be somewhat worse.

\begin{figure}[htb]
    \centering
    \includegraphics[width=\cmsFigWidth]{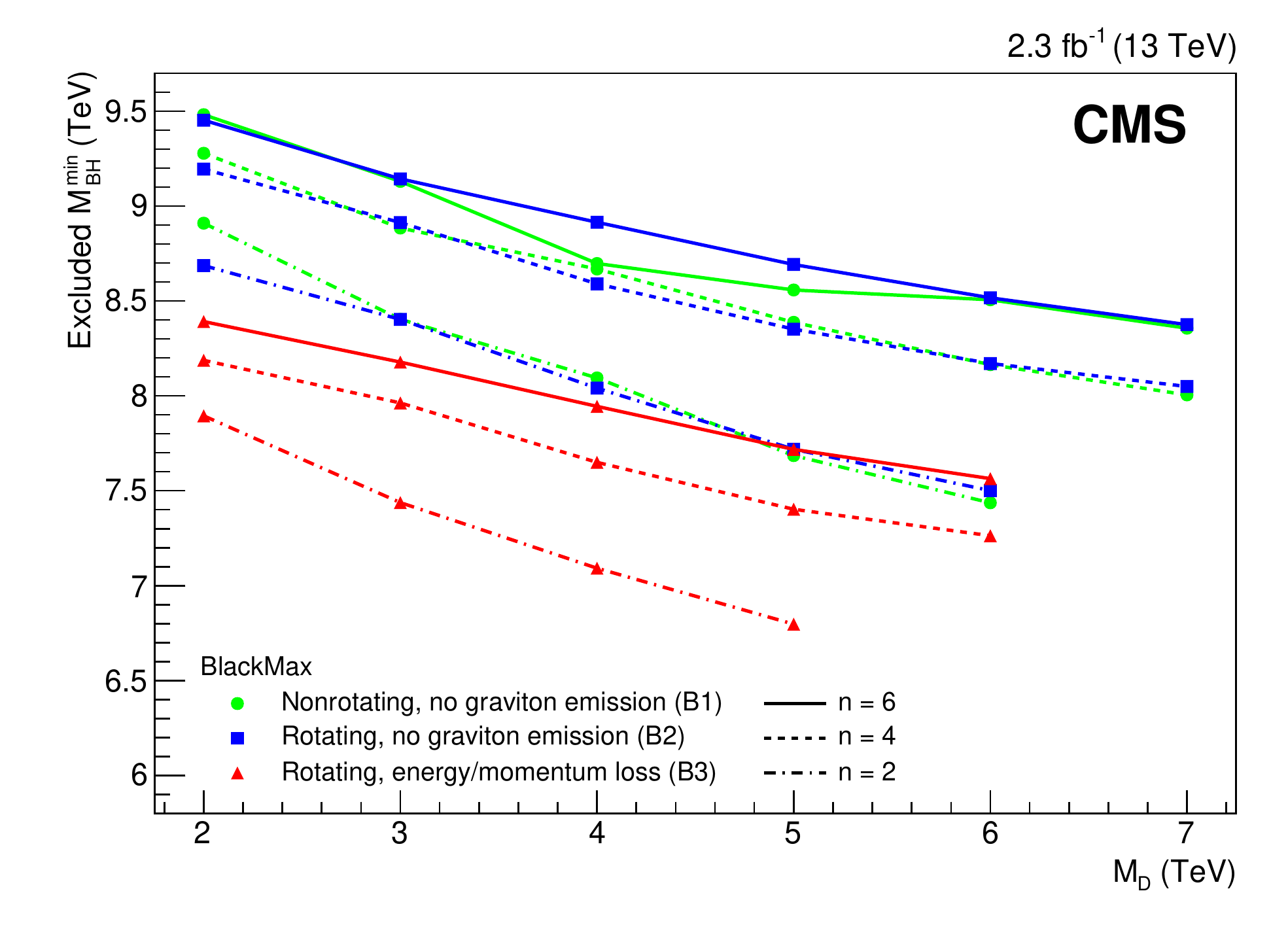}
    \caption{The 95\% CL  lower limits on the minimum semiclassical black hole mass as a function of the Planck scale \MD, for several benchmark models generated with {\BLACKMAX}: nonrotating and rotating black holes without graviton emission and rotating black holes with energy-momentum loss.}
    \label{fig:BHlimit}
\end{figure}

\begin{figure}[htb]
    \centering
    \includegraphics[width=\cmsFigWidth]{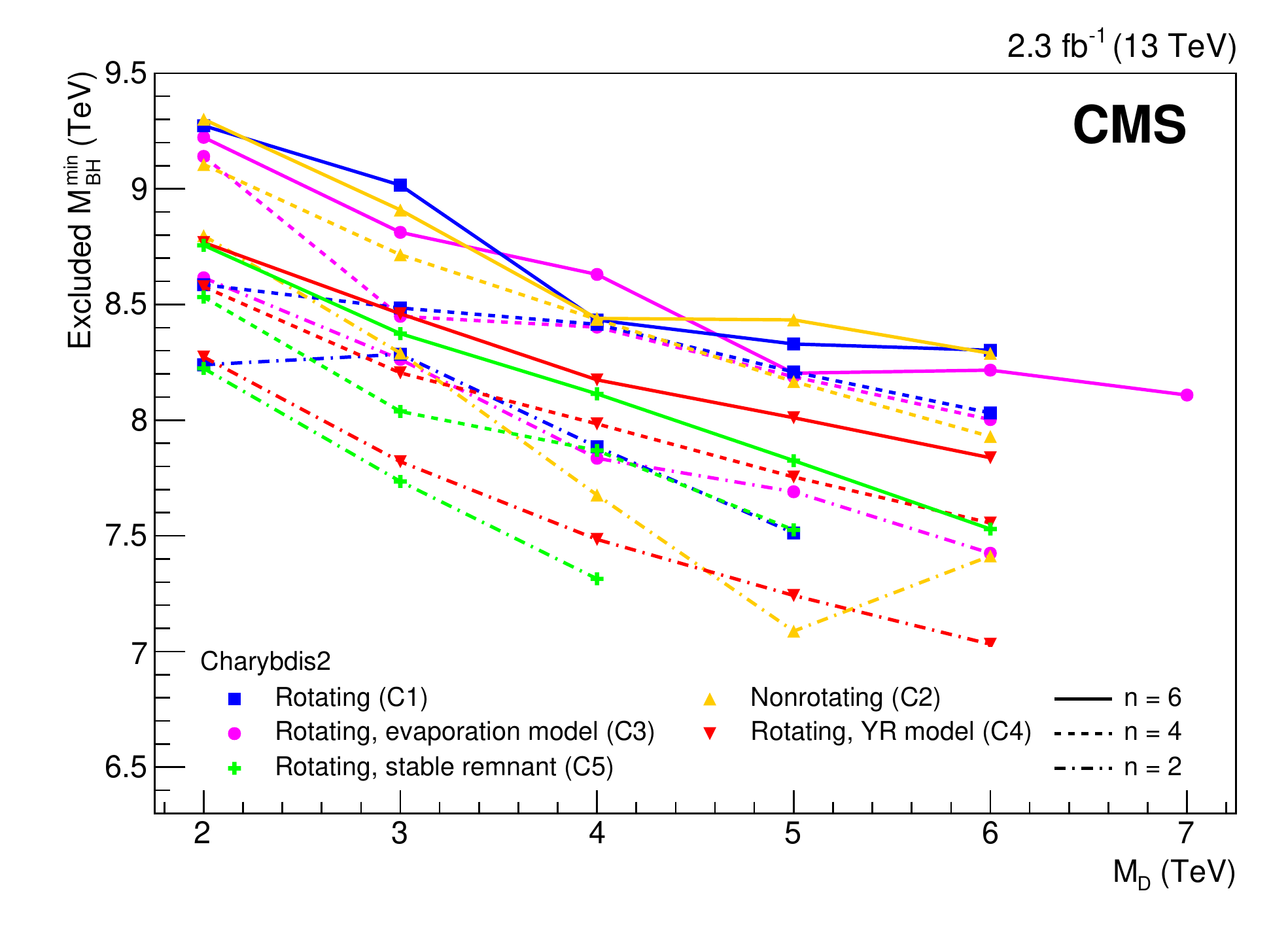}
    \caption{The 95\% CL lower limits on minimum semiclassical black hole mass as a function of the Planck scale \MD, for several benchmark models generated with {\CHARYBDIS2}: rotating and nonrotating black holes, rotating black holes with an alternative evaporation model, rotating black holes with Yoshino--Rychkov bounds, and rotating black holes with a stable remnant.}
    \label{fig:Limit_Charybdis}
\end{figure}

Model-specific limits span the entire set of models discussed in Section~\ref{sec:signal}. The limits on semiclassical BHs in all cases, except for the model with stable remnant (C5), come from the  optimized values of $N^\text{min}$ equal to 9 or 10 for low BH masses and from lower values of $N^\text{min}$ for high BH masses, thanks to smaller backgrounds at high values of \st. In Fig.~\ref{fig:BHlimit}, we show the observed lower limits on the semiclassical BH mass at 95\% CL for {\BLACKMAX} signal samples. In these models, we exclude minimum BHs masses below 7.0--9.5\TeV, for a large set of model parameters. Similar limits for models generated with the {\CHARYBDIS2} generator are shown in Fig.~\ref{fig:Limit_Charybdis}. Considering that an independent optimization was used for each model, and for each $\MD$ and $n$ point, these limits are in general agreement with those obtained using {\BLACKMAX} for analogous models. The limits significantly extend those from LHC Run 1~\cite{CMSBH3}, which only reach 6.5\TeV.

In the case of QBHs and the case of semiclassical BHs with a stable remnant, where the number of produced particles is small as the massive stable remnant escapes detection, the best sensitivity is achieved for $N^\text{min} = 2$. This sample partially overlaps with the $N=2$ data set used for deriving the background template. Nevertheless, even in this case we still can use the background estimate based on the $N=2$ spectrum, because the background is determined using only the low-\st range of the spectrum (1.4--2.4\TeV), where the  signal has already been excluded up to $\approx$5\TeV by the 8\TeV analysis~\cite{CMSBH3}. Thus the potential signal contamination of the \st range used in deriving the background shape and normalization is negligible.

\begin{figure}[htb]
    \centering
    \includegraphics[width=\cmsFigWidth]{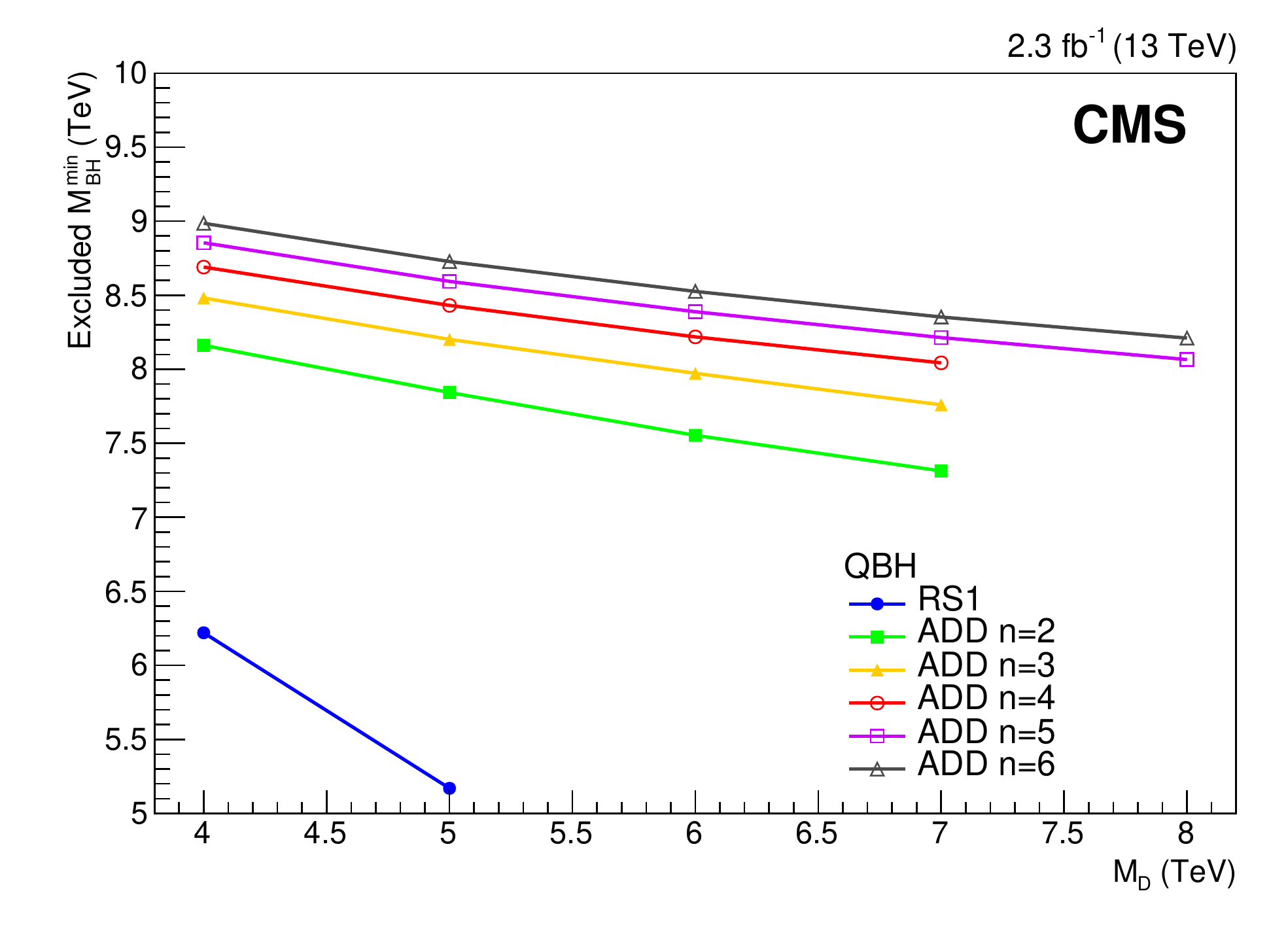}
    \caption{The 95\% CL lower limits on minimum quantum black hole mass as a function of the Planck scale \MD, for several benchmark models. The blue (lower) line corresponds to quantum black holes in the RS1 model; while the other lines correspond to the ADD model for the number of extra dimensions $n = 2,\ldots,6$.}
    \label{fig:QBHLimit}
\end{figure}

The lower limits on the minimum QBH mass are shown in Fig.~\ref{fig:QBHLimit} and span the 7.3--9.0\TeV range for the ADD ($n > 2$) and 5.1--6.2\TeV range for the RS1 ($n=1$) QBHs. Again, these limits extend significantly those obtained in the LHC Run 1 (4.6--6.2\TeV).

\begin{figure}[htb]
    \centering
    \includegraphics[width=\cmsFigWidth]{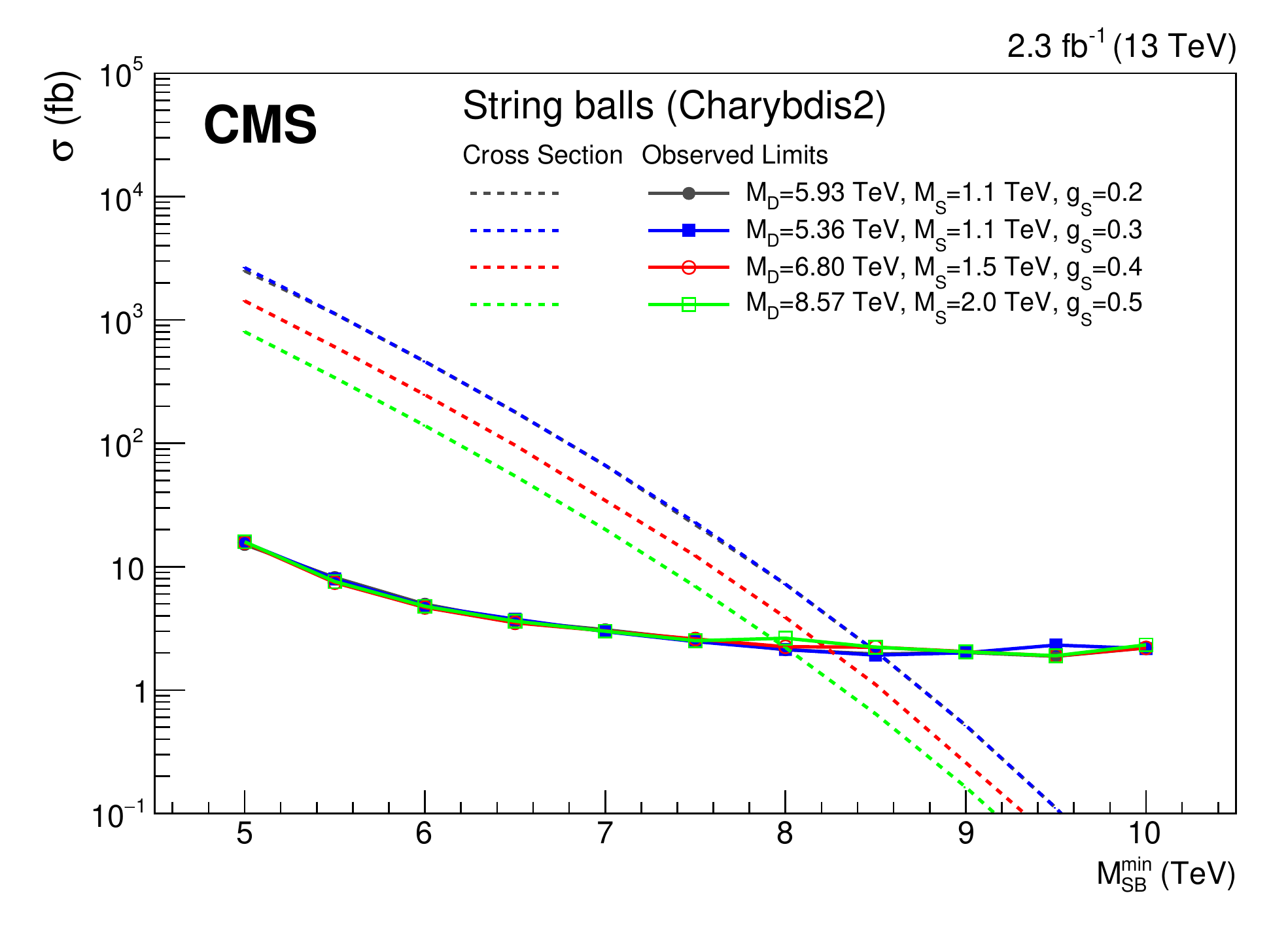}
    \caption{The 95\% CL upper limits on the cross section for the production of string balls and the corresponding theoretical cross sections.
 The solid colored lines correspond to the observed cross section limits; the dashed colored curves correspond to theoretical cross sections. }
    \label{fig:Limit_StringBalls}
\end{figure}

Finally, for the case of the SBs, the mass exclusion limits shown in Fig.~\ref{fig:Limit_StringBalls} reach 8.0--8.5\TeV. Most of these limits correspond to the saturated string ball regime, $\MS/\gs < M_\mathrm{SB} < \MS/\gs^2$. The transitions between the saturated SB and BH regimes are not clearly visible in theoretical cross section curves, as the parton-level cross section that exhibits this transition as a change in slope is significantly modified by the rapidly falling PDFs.

\section{Summary\label{s:summary}}

We have conducted a search for new physics in multiparticle final states in a data sample of proton-proton collisions at $\sqrt{s} = 13$\TeV collected with the CMS detector, corresponding to an integrated luminosity of 2.3\fbinv. The discriminating variable between signal and the dominant QCD multijet background is the scalar sum of the transverse energies of all reconstructed objects in the event, \st. The shape of the \st distribution in low-multiplicity data is used to predict the QCD multijet background in high-multiplicity signal regions. No significant excess of events over the standard model expectation is observed in any of the analyzed final-state multiplicities. Comparing the \st distribution in data with that from the background prediction, we set model-independent upper limits at 95\% confidence level on the product of the cross section and
the acceptance for hypothetical signals. In addition, we set limits on various theoretical black hole and string ball
models, including models of rotating and nonrotating black holes and quantum black holes. In all cases the exclusions represent significant improvements over the limits achieved in Run 1 of the LHC.

\begin{acknowledgments}
We congratulate our colleagues in the CERN accelerator departments for the excellent performance of the LHC and thank the technical and administrative staffs at CERN and at other CMS institutes for their contributions to the success of the CMS effort. In addition, we gratefully acknowledge the computing centers and personnel of the Worldwide LHC Computing Grid for delivering so effectively the computing infrastructure essential to our analyses. Finally, we acknowledge the enduring support for the construction and operation of the LHC and the CMS detector provided by the following funding agencies: BMWFW and FWF (Austria); FNRS and FWO (Belgium); CNPq, CAPES, FAPERJ, and FAPESP (Brazil); MES (Bulgaria); CERN; CAS, MoST, and NSFC (China); COLCIENCIAS (Colombia); MSES and CSF (Croatia); RPF (Cyprus); SENESCYT (Ecuador); MoER, ERC IUT, and ERDF (Estonia); Academy of Finland, MEC, and HIP (Finland); CEA and CNRS/IN2P3 (France); BMBF, DFG, and HGF (Germany); GSRT (Greece); OTKA and NIH (Hungary); DAE and DST (India); IPM (Iran); SFI (Ireland); INFN (Italy); MSIP and NRF (Republic of Korea); LAS (Lithuania); MOE and UM (Malaysia); BUAP, CINVESTAV, CONACYT, LNS, SEP, and UASLP-FAI (Mexico); MBIE (New Zealand); PAEC (Pakistan); MSHE and NSC (Poland); FCT (Portugal); JINR (Dubna); MON, RosAtom, RAS, RFBR and RAEP (Russia); MESTD (Serbia); SEIDI, CPAN, PCTI and FEDER (Spain); Swiss Funding Agencies (Switzerland); MST (Taipei); ThEPCenter, IPST, STAR, and NSTDA (Thailand); TUBITAK and TAEK (Turkey); NASU and SFFR (Ukraine); STFC (United Kingdom); DOE and NSF (USA).

\hyphenation{Rachada-pisek} Individuals have received support from the Marie-Curie program and the European Research Council and EPLANET (European Union); the Leventis Foundation; the A. P. Sloan Foundation; the Alexander von Humboldt Foundation; the Belgian Federal Science Policy Office; the Fonds pour la Formation \`a la Recherche dans l'Industrie et dans l'Agriculture (FRIA-Belgium); the Agentschap voor Innovatie door Wetenschap en Technologie (IWT-Belgium); the Ministry of Education, Youth and Sports (MEYS) of the Czech Republic; the Council of Science and Industrial Research, India; the HOMING PLUS program of the Foundation for Polish Science, cofinanced from European Union, Regional Development Fund, the Mobility Plus program of the Ministry of Science and Higher Education, the National Science Center (Poland), contracts Harmonia 2014/14/M/ST2/00428, Opus 2014/13/B/ST2/02543, 2014/15/B/ST2/03998, and 2015/19/B/ST2/02861, Sonata-bis 2012/07/E/ST2/01406; the National Priorities Research Program by Qatar National Research Fund; the Programa Clar\'in-COFUND del Principado de Asturias; the Thalis and Aristeia programs cofinanced by EU-ESF and the Greek NSRF; the Rachadapisek Sompot Fund for Postdoctoral Fellowship, Chulalongkorn University and the Chulalongkorn Academic into Its 2nd Century Project Advancement Project (Thailand); and the Welch Foundation, contract C-1845.
\end{acknowledgments}

\bibliography{auto_generated}

\cleardoublepage \appendix\section{The CMS Collaboration \label{app:collab}}\begin{sloppypar}\hyphenpenalty=5000\widowpenalty=500\clubpenalty=5000\textbf{Yerevan Physics Institute,  Yerevan,  Armenia}\\*[0pt]
A.M.~Sirunyan, A.~Tumasyan
\vskip\cmsinstskip
\textbf{Institut f\"{u}r Hochenergiephysik,  Wien,  Austria}\\*[0pt]
W.~Adam, E.~Asilar, T.~Bergauer, J.~Brandstetter, E.~Brondolin, M.~Dragicevic, J.~Er\"{o}, M.~Flechl, M.~Friedl, R.~Fr\"{u}hwirth\cmsAuthorMark{1}, V.M.~Ghete, C.~Hartl, N.~H\"{o}rmann, J.~Hrubec, M.~Jeitler\cmsAuthorMark{1}, A.~K\"{o}nig, I.~Kr\"{a}tschmer, D.~Liko, T.~Matsushita, I.~Mikulec, D.~Rabady, N.~Rad, B.~Rahbaran, H.~Rohringer, J.~Schieck\cmsAuthorMark{1}, J.~Strauss, W.~Waltenberger, C.-E.~Wulz\cmsAuthorMark{1}
% ...
\vskip\cmsinstskip
% ...
1:~~Also at Vienna University of Technology, Vienna, Austria\\
% ...

\end{sloppypar}
\end{document}